\documentclass[10pt,journal,twoside,final]{IEEEtran}

\usepackage{diagbox}
\usepackage{rotating}
\usepackage{graphicx,subfig}
\usepackage{cite}
\usepackage{enumitem}
\usepackage{booktabs} 
\usepackage{color}
\usepackage{bigstrut}
\usepackage{multirow}

\usepackage{amsmath, amsthm, amssymb}
\usepackage{url}
\usepackage{array}
\newcolumntype{L}[1]{>{\raggedright\let\newline\\\arraybackslash\hspace{0pt}}m{#1}}
\newcolumntype{C}[1]{>{\centering\let\newline\\\arraybackslash\hspace{0pt}}m{#1}}
\newcolumntype{R}[1]{>{\raggedleft\let\newline\\\arraybackslash\hspace{0pt}}m{#1}}
\usepackage{amssymb}

\newcommand{\etal}{\textit{et al. }}
\begin{document}

% From Heuristics to Learned Embedding
\title{Secrets in Preamble: Practical and Robust PHY Information Hiding for Wi-Fi}

% \title{From Natural to Digital: Towards Reliable RF Fingerprinting with Quantifiable User Capacity}

% \title{From Passive to Active: Towards Reliable RF Fingerprinting with Quantifiable User Capacity}

\author{Rui~Pan,
    Guanxiong~Shen,~\IEEEmembership{Member,~IEEE},
    Linning~Peng,~\IEEEmembership{Member,~IEEE},
    Xinyu~Qi,
	Junqing~Zhang,~\IEEEmembership{Senior Member,~IEEE}
    
	\thanks{Manuscript received xxx; revised xxx; accepted xxx. Date of publication xxx; date of current version xxx. The work of 
    The review of this paper was coordinated by xxx. 
    \textit{(Corresponding author: Guanxiong Shen.)}}
    \thanks{R.~Pan, G.~Shen and L.~Peng are with the School of Cyber Science and Engineering, Southeast University, Nanjing, China. (email: 230259649@seu.edu.cn; gxshen@seu.edu.cn; pengln@seu.edu.cn)}
    \thanks{X.~Qi is with the School of Communications and Information Engineering, Nanjing University of Posts and Telecommunications, Nanjing 210003, China, and also with the Jiangsu Key Laboratory of Intelligent Information Processing and Communication Technology, Nanjing University of Posts and Telecommunications, Nanjing 210003, China. (e-mail: xyqi@njupt.edu.cn)}
	\thanks{J.~Zhang is with the School of Computer Science and Informatics, University of Liverpool, Liverpool, L69 3DR, United Kingdom. (email: Junqing.Zhang@liverpool.ac.uk)}
    
    % \thanks{A.~Hu is with the National Mobile Communications Research Laboratory, Southeast University, Nanjing 210096, China, and also with Purple Mountain Laboratories, Nanjing 211111, China. (email: aqhu@seu.edu.cn)}
    
	\thanks{Color versions of one or more of the figures in this paper are available online at http://ieeexplore.ieee.org.}
	\thanks{Digital Object Identifier xxx}	
}

% \author{
% 	\IEEEauthorblockN{
% 		Author 1,
% 		Author 2,
% 	    Author 3,
% 		Author 4,
%         and Author 5
% 	}

% }

\maketitle

\begin{abstract}

Physical layer (PHY) information hiding refers to the concealment of secrets within PHY waveforms without compromising the integrity of the primary communication link, thereby enabling critical applications such as covert communication, PHY steganography, and watermarking. The widespread deployment of Wi-Fi has motivated substantial research interest in hiding secrets within Wi-Fi waveforms. However, existing Wi-Fi information hiding approaches suffer from two limitations: restricted steganographic capacity and insufficient robustness against environmental variations, both of which severely constrain practical deployment.
This work proposes to embed secrets within the Wi-Fi legacy short training field (L-STF). The designed method preserves the autocorrelation properties of Wi-Fi L-STF, ensuring that packet detection and coarse frequency offset compensation functionalities remain uncompromised. The secret embedding and extraction processes employ a pair of neural networks designed to learn optimal embedding rules and maximize steganographic capacity. We developed a prototype system and conducted comprehensive experiments across both indoor and outdoor environments. Experimental results show that our method can reliably embed 46 secret bits within Wi-Fi packets across diverse channel conditions, achieving significantly enhanced steganographic capacity and environmental robustness. 
The hardware prototype study validates the feasibility of the proposed Wi-Fi information hiding scheme.
% Experimental results demonstrate that the secret can be reliably extracted with a secret bit error rate (SBER) less than 0.17, validating the feasibility and effectiveness of the proposed Wi-Fi information hiding scheme. 
% The code and dataset will be released publicly upon publication.

\end{abstract}


\section{Introduction}

Physical layer (PHY) information hiding embeds secret information into wireless signals by deliberately introducing subtle and imperceptible distortions into PHY waveforms. This technique enables a range of critical applications, including PHY steganography for covert communication~\cite{dutta2012secret, rahbari2017exploiting, schulz2018shadow, d2019hiding, zheng2019multi, zhang2021joint, wang2021covert, zhang2021adaptive, jiao2021openwifi, hou2022cloaklora, du2022performance, wei2023wise, liu2024exploiting} and PHY watermarking for transmitter authentication~\cite{verma2015physical, zhang2020physical, an2021tag, zhang2023tag, tan2025asynchronous}. Hence, it has attracted significant research attention.
Within this paradigm, secret information is embedded into ubiquitous wireless signals, such as Wi-Fi~\cite{rahbari2017exploiting, schulz2018shadow}, LoRa~\cite{hou2022cloaklora}, satellite communications~\cite{chen2019intelligent}, and cellular networks~\cite{bonati2021stealte}, through subtle PHY distortions that do not compromise signal integrity~\cite{grabska2013steganography, yang2017nicscatter, li2019achieving, zheng2019multi, wang2021covert, zhang2021joint, du2022performance, zhang2023tag, liu2024exploiting}. Consequently, external observers can receive and demodulate the conventional payload data without detecting the presence of concealed secrets. In contrast, only authorized receivers equipped with appropriate secret extraction mechanisms are capable of recovering the embedded information.
PHY information hiding operates at the most fundamental layer of wireless communication systems, offering distinct advantages over higher-layer steganographic techniques~\cite{chen2023covert}. 
% This approach remains virtually undetectable to ordinary wardens, primarily because most commercial wireless devices do not provide direct access to PHY information~\cite{jiang2021eliminating}, thereby creating an inherent hardware barrier against unauthorized detection and eavesdropping.

Given that Wi-Fi has become the \textit{de facto} wireless local area network (WLAN) standard, Wi-Fi infrastructures are ubiquitous in daily life, making the concealment of secrets within Wi-Fi transmissions attract considerable research interests. Some prior studies~\cite{schulz2018shadow, jiao2021openwifi} propose embedding secrets within transmitted Wi-Fi signals through artificial filtering, where the receiver can extract secrets by analyzing the estimated channel state information (CSI). However, the authors in~\cite{jiao2021openwifi} claim that such an approach is valid only in static environments, and the Shadow Wi-Fi system proposed in~\cite{schulz2018shadow} can only embed a 2-bit secret per packet, demonstrating limited steganographic capacity.
Another category of approaches embeds secrets by manipulating the constellation diagram. For example, the authors in~\cite{dutta2012secret} propose a `dirty' constellation scheme to hide secrets. D'Oro~\etal further designed a pseudo-noise asymmetric shift-keying (PN-ASK) modulation for covert Wi-Fi systems, concealing additional information in the symbol amplitude~\cite{d2019hiding}. However, these constellation-manipulating schemes inevitably degrade the decoding performance of the primary Wi-Fi link, resulting in increased error rates when decoding the payload. Therefore, it is necessary to explore approaches that can achieve satisfactory steganographic capacity without compromising the primary Wi-Fi channel.

Hiding secrets within Wi-Fi transmissions presents several significant technical challenges~\cite{jiang2024physical}. First, any artificial distortion applied to the transmitted signal must not interfere with standard Wi-Fi demodulation procedures, ensuring that the primary communication link remains uncompromised and appears normal to wardens. Second, since Wi-Fi operates as a wideband communication system with bandwidths of at least 20 MHz, the embedded secrets must be robust enough to survive various wireless propagation effects such as multipath fading and noise. Third, it is challenging to manually design an optimal secret embedding rule that achieves maximum steganographic capacity while simultaneously satisfying the aforementioned dual requirements. These interconnected issues require innovative strategies that maintain a careful balance between the efficiency of steganographic methods and the real-world limitations faced by Wi-Fi systems.

In this work, we propose a method to embed secret information into the Wi-Fi legacy short training field (L-STF) using a neural network (NN), with the receiver employing a paired NN for extracting the embedded secrets. This pair of NNs adopts an encoder-decoder architecture and undergoes joint training on simulation platforms. To enhance robustness against channel variations, we integrate a TGn channel simulator into the training process and apply L-STF equalization to mitigate channel effects. Collectively, these solutions enable a \textit{`train once in simulation, deploy anywhere in the real world'} capability. A prototype for Wi-Fi information hiding is developed using commercial software-defined radio (SDR) platforms, and experimental evaluations are conducted in both indoor and outdoor scenarios. The detailed contributions are summarized as follows.
\begin{itemize}
    \item We propose a PHY information hiding system that can embed secrets within the Wi-Fi L-STF. The periodic repeating property of L-STF is preserved, ensuring that packet detection and coarse frequency offset compensation functionalities remain unaffected.
    
    \item The secret embedding and extraction are realized through a pair of NNs. Specifically, the secret embedder and extractor are jointly trained using an encoder-decoder architecture. The training is conducted on a simulation platform, and the trained models are then securely distributed to the transmitter and receiver.

    \item Two strategies are employed to enhance the system's robustness against channel variations. First, a TGn channel simulator is incorporated during training to improve the generalization ability of NNs. Second, equalization is applied not only to the payload but also to the Wi-Fi L-STF to mitigate channel distortion, i.e., L-STF equalization. Notably, the model trained in simulation can be directly deployed on real-world devices without any fine-tuning.
    
    \item We created a prototype using commodity SDR and carried out experiments in both indoor and outdoor environments. The experiments demonstrate that the proposed Wi-Fi information hiding system can effectively embed a 46-bit secret message into the L-STF without affecting the payload demodulation process.

\end{itemize}

\textbf{The reproducible code, simulation tools, and experimental dataset will be made available to the research community upon acceptance of the manuscript.}

The remainder of this paper is organized as follows. Section~\ref{sec:related_work} summarizes related work. Section~\ref{sec:system_overview} presents an overview of the proposed system and discusses the core technical challenges. Sections~\ref{sec:alice} and~\ref{sec:bob} detail the design of the secret embedding process at the transmitter (Alice) and the secret extraction process at the receiver (Bob), respectively. Section~\ref{sec:training} describes the end-to-end training methodology for the NN pair, including the integrated channel simulation and the composite loss function. The performance of our scheme is then evaluated through comprehensive simulations in Section~\ref{sec:simulation_results} and real-world over-the-air experiments in Section~\ref{sec:experimental_results}. Section~\ref{sec:discussion} analyzes the security, complexity, and compatibility of the proposed scheme. Section~\ref{sec:conclusion} concludes the paper.

% \clearpage

% In fact, several RF fingerprinting studies conduct evaluation on a dataset containing over 10,000 Wi-Fi devices, and the identification accuracy is only 40\%. 
% Since the dataset is not public, it is difficult to tell can the identification accuracy be improved to a satisfactory performance. 

\section{Related Work}\label{sec:related_work}

This section reviews information hiding in conventional digital media and covert communication at the wireless PHY layer.

\textbf{Information hiding} refers to concealing sensitive information within ordinary media. Conventional methods rely on handcrafted rules. A representative example is the least significant bits (LSB) method~\cite{van1994digital}, which substitutes the least significant bit planes of the cover. The work in~\cite{hopper2002provably} established the equivalence between one-way functions and secure steganographic protocols. Within this discipline, digital watermarking for copyright protection and steganography for covert communication are two representative applications. Reversible embedding restores the cover after the secret is read. Capacity is increased by neighbour-mean interpolation~\cite{jana2022new}, difference expansion~\cite{mukherjee2019novel}, or spreading the secret over a pair of images~\cite{lu2023dual}, while robustness is pursued with a weighted matrix and local binary patterns~\cite{pal2021secure} or Hamming coding on block-truncation-coded images~\cite{datta2024robust}. These methods assume a cover that is stored and read back exactly, which a wireless channel does not provide.

Deep learning has shifted this field from handcrafted embedding rules to strategies learned through loss functions. Zhu~\etal~\cite{zhu2018hidden} introduced the first end-to-end framework for hiding secrets in images, and Singh~\etal~\cite{singh2023riwt} proposed an RIWT residual network for medical images. Subsequent work includes secret concealment with all-optical NNs~\cite{bai2024information}, watermarking of deep learning models~\cite{adi2018turning}, and hiding information in NN outputs~\cite{jois2024pulsar}. Beyond images, Wu~\etal~\cite{wu2024generative} embed text through the user interface of large language models (LLMs) without access to model internals, and Geleta~\etal~\cite{geleta2022pixinwav} embed images into audio spectrograms. These approaches are not directly transferable to wireless signals, which are subject to multipath fading, ambient noise, and hardware impairments, and thus motivate information hiding at the PHY layer.
% , such as text~\cite{wu2024generative, silva2024look}, audio~\cite{geleta2022pixinwav, pang2024fremax}, and network traffic~\cite{fathi2020gan, gong2022surakav}.

\textbf{Covert communication} is essential in highly sensitive applications, where conventional encryption alone is insufficient to satisfy the security demands~\cite{chen2023covert}. In these scenarios, it is equally critical to obscure the communication process, ensuring that transmissions remain undetectable to potential adversaries.

Prior research has explored various strategies to achieve covert communication through PHY layer design. For instance, Zheng~\etal investigated multi-antenna-aided covert communication in both centralized and distributed antenna systems~\cite{zheng2019multi}. Du~\etal explored covert communication for multi-antenna unmanned aerial vehicle systems aided by jamming, deriving closed-form expressions for signal to interference plus noise ratio (SINR) distributions, detection error probabilities, and covert rates under composite fading and shadowing channels~\cite{du2022performance}. Beyond these antenna-based approaches, additional studies leveraged millimeter-wave (mmWave)\cite{wang2021covert,zhang2021joint} and reconfigurable intelligent surface (RIS) techniques\cite{zhang2023tag, liu2024exploiting} to enhance covertness in wireless systems.

However, most of these approaches are largely limited to theoretical analysis.
Emerging studies have begun to explore the embedding of secret information within existing PHY transmissions, in line with the concept of information hiding. For instance, D'Oro~\etal introduced PN-ASK modulation into phase-modulated Wi-Fi systems, enabling the concealment of extra information through symbol amplitude manipulation~\cite{d2019hiding}. Similarly, Hou~\etal proposed CloakLoRa~\cite{hou2022cloaklora}, which modulates the amplitude of frequency-modulated LoRa signals without altering the carrier frequency, thereby establishing a covert channel for secret transmission. More recently, Wei and Saha proposed a wireless steganography scheme that encodes secrets into signals statistically mimicking transmitter hardware noise, enabling covert embedding within arbitrary orthogonal frequency division multiplexing (OFDM) waveforms~\cite{wei2023wise}.
These approaches demonstrate the capability to embed secrets within conventional Wi-Fi, ZigBee, and LoRa PHY waveforms, thus facilitating practical covert communication implementations.

\section{System Overview and Challenges}\label{sec:system_overview}
This section first presents the overall workflow of the proposed PHY information hiding system, detailing how secret embedding/extraction aligns with standard Wi-Fi PHY operations. We then discuss the core technical challenges specific to Wi-Fi-based PHY information hiding.

\subsection{System Overview}
% \begin{figure*}[!t]
%     \centering
%     \includegraphics[width=6.4in]{figures/system_overview.pdf}
%     \caption{Overview of the Wi-Fi information hiding system. \red{hat_h arrow should be upsidedown}}
%     \label{fig:system_overview}
% \end{figure*}
\begin{figure}[!t]
    \centering
    \includegraphics[width=3.4in]{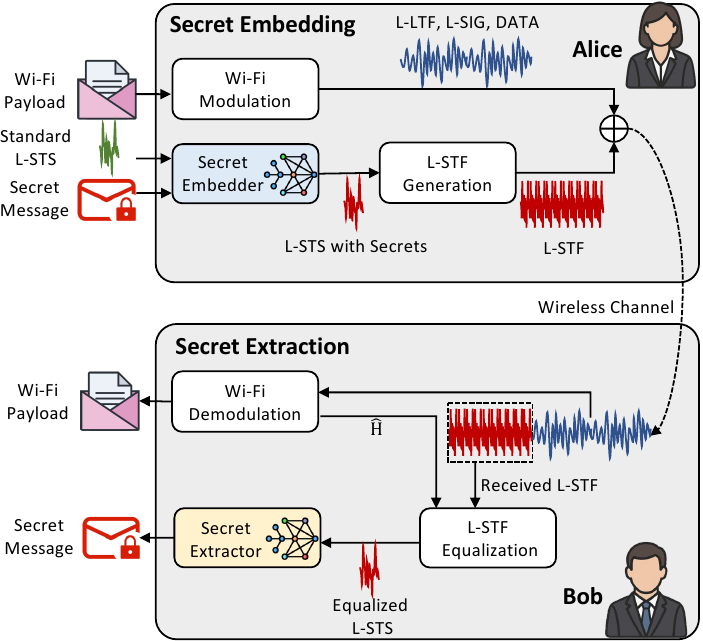}
    \caption{Overview of the Wi-Fi information hiding system.}
    \label{fig:system_overview}
\end{figure}

Fig.~\ref{fig:system_overview} illustrates the overall architecture of the proposed PHY information hiding system. Two legitimate parties, Alice and Bob, establish a standard Wi-Fi communication link. Alice embeds secret messages into a complete Wi-Fi packet waveform by introducing subtle and imperceptible distortions using a pre-trained neural network (NN), referred to as the \emph{secret embedder}. At the receiver, Bob employs a paired NN, termed the \emph{secret extractor}, to recover the hidden secrets from the received packet waveform, while still performing standard demodulation to decode the Wi-Fi payload. The operations at Alice and Bob are summarized as follows.

\textbf{Secret Embedding at Alice:}  
To convey a secret message through a covert channel, Alice first embeds the secret into the legacy short training field (L-STF) using the NN-based secret embedder. The embedder takes as input the secret message and the standard L-STF, and outputs a slightly distorted L-STF carrying the hidden information. Meanwhile, the remaining Wi-Fi fields are generated following the IEEE 802.11 standard. Specifically, the legacy long training field (L-LTF) is produced for channel estimation, the legacy signal field (L-SIG) encodes the transmission parameters (e.g., rate and payload length), and the payload bits are modulated into the DATA field. These standard components, referred to as the \emph{Unmodified Region} in Fig.~\ref{fig:format}, are concatenated with the distorted L-STF and then transmitted over the air through the RF front-end. The detailed secret embedding process is described in Section~\ref{sec:alice}.

\textbf{Secret Extraction at Bob:}  
Upon receiving the packet waveform, Bob first performs standard Wi-Fi demodulation on the unmodified fields (L-LTF, L-SIG, and DATA) to successfully recover the payload. During this process, the receiver also obtains the channel state information (CSI) from the L-LTF, which is normally used for payload equalization. In the proposed system, this CSI is further reused to equalize the L-STF region, referred to as \emph{L-STF equalization}, in order to mitigate channel-induced distortions on the secret-bearing preamble. Finally, the equalized L-STF is fed into the NN-based secret extractor, which decodes the embedded secret message. The detailed extraction process is presented in Section~\ref{sec:bob}.

\textbf{Training of Neural Networks:}  
The secret embedder and secret extractor are jointly trained in an end-to-end encoder–decoder architecture. Training is conducted in a controlled simulation environment that emulates realistic wireless channels, enabling the networks to learn robust embedding and extraction rules while keeping the waveform distortions imperceptible. After training, the model weights are securely distributed to Alice and Bob, ensuring that the NNs remain confidential and inaccessible to potential adversaries. The training methodology is presented in Section~\ref{sec:training}.

\subsection{Motivation, Challenges and Solutions}\label{sec:challenges}

This work aims to design a PHY information hiding scheme for Wi-Fi. Specifically, the objective is to introduce imperceptible and subtle distortion into the outgoing Wi-Fi waveform while maintaining the packet payload demodulation unaffected. The challenges are summarized as follows:
\begin{itemize}
    \item[(a)] \textbf{Preserving Standard Wi-Fi Demodulation Performance:} A fundamental challenge in designing PHY information hiding systems is ensuring that the embedded distortion does not interfere with normal Wi-Fi reception and demodulation processes, thereby making the Wi-Fi transmissions appear normal to wardens.
    
    \item[(b)] \textbf{Robustness to Channel Variations:} The second challenge lies in ensuring that embedded secrets remain extractable after experiencing over-the-air propagation. The system must be sufficiently robust to survive realistic multipath wireless channel conditions.

    \item[(c)] \textbf{Sufficient Steganographic Capacity:} Thirdly, it is challenging to manually design a secret embedding rule that satisfies the above requirements while achieving optimal steganographic capacity.
\end{itemize}

In this study, we propose a series of solutions to address these challenges. For challenge (a), we apply identical PHY distortions to all ten short training symbols (STSs). Since the periodic repetition property remains uncompromised, the packet detection and synchronization functionalities are preserved. Moreover, normal packet demodulation remains unaffected as the payload waveform is kept unchanged. Regarding challenge (b), we apply channel equalization to both the Wi-Fi payload and the L-STF segment, which significantly mitigates channel effects. For challenge (c), we utilize a pair of NNs to learn optimal secret embedding and extraction rules. We define appropriate loss functions and incorporate a TGn channel simulator to enhance model robustness.

Information hiding in images and audio is a well-developed field.
It has advanced substantially with the advent of deep neural networks~\cite{zhu2018hidden}.
However, the settings between image/audio and radio waveforms are not equivalent.
Image and audio steganography need only keep the cover acceptable to human vision or hearing.
A Wi-Fi waveform must additionally remain a functional communication signal.
The embedding must not interfere with standard Wi-Fi packet detection, synchronization, or payload demodulation, which corresponds to challenge~(a).
In addition, in contrast to an image file that can be conveyed without distortion, the secret must also remain extractable after wireless propagation, which corresponds to challenge~(b).
Addressing these two radio-specific requirements is the key contribution of this work.

% \blue{The security of the proposed scheme is analyzed in Section~\ref{sec:security_analysis}.

\section{Secrets Embedding at Alice}\label{sec:alice}
As shown in Fig.~\ref{fig:system_overview}, Alice embeds secrets within the packet waveform. This section first justifies the selection of L-STF as the secret embedding field, then presents the design of the NN-based secret embedder, and finally provides a visualization of the resulting distorted L-STF to demonstrate the imperceptibility.

\subsection{Selection of Secret Embedding Field}\label{sec:embedding_field}

In the proposed Wi-Fi information hiding scheme, secret messages are embedded as subtle PHY distortions within the L-STF region, while all other fields remain unchanged. Specifically, we apply distortion to a single STS and repeat it ten times to construct the complete L-STF region. This section explains the rationale behind this design choice by examining the Wi-Fi packet structure and the functional characteristics of L-STF.

\begin{figure}[!t]
	\centering
    \includegraphics[width=3.4in]{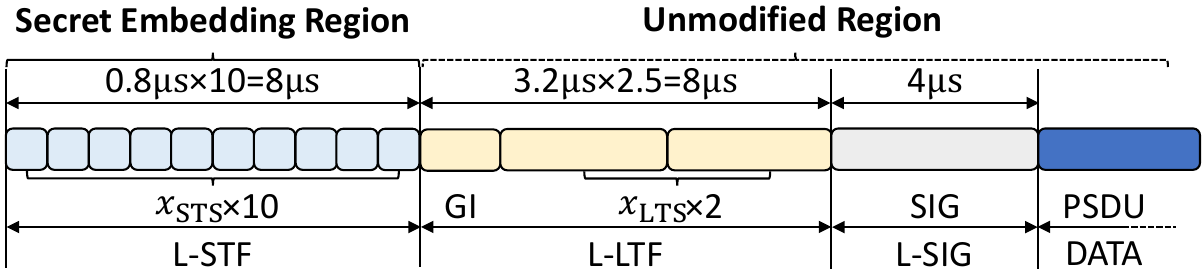}
	\caption{PHY format of a 20~MHz non-HT OFDM Wi-Fi transmission.}
	\label{fig:format}
\end{figure}

\subsubsection{Wi-Fi Packet Structure}
As depicted in Fig.~\ref{fig:format}, the L-STF consists of ten identical STSs and facilitates packet detection and coarse carrier frequency offset (CFO) compensation. Following the L-STF, the legacy long training field (L-LTF) comprises a guard interval (GI) and two identical long training symbols (LTSs), where the GI is defined as the latter half of an LTS. The L-LTF enables fine time synchronization, precise CFO compensation, and CSI estimation. The legacy signal (L-SIG) field conveys modulation rate, frame length, parity check information, and tail bits. Finally, the DATA field carries the PHY service data unit (PSDU), which encapsulates the MAC frame payload.

Understanding this structure reveals why L-STF is suitable for secret embedding. Introducing distortions into the L-LTF region would lead to channel estimation errors, thereby impairing accurate demodulation of the DATA field and potentially interfering with the primary Wi-Fi communication link. Similarly, modifying the L-SIG or DATA fields would directly compromise the integrity of the transmitted payload. \textbf{In contrast, the L-STF's core functionalities can be preserved as long as its periodic structure is maintained}, making it the ideal candidate for secret embedding without degrading standard Wi-Fi demodulation performance.

\subsubsection{Function of Wi-Fi L-STF}\label{sec:function_lstf}

\begin{figure}[!t]
	\centering
    \subfloat[]{\includegraphics[width=1.7in]{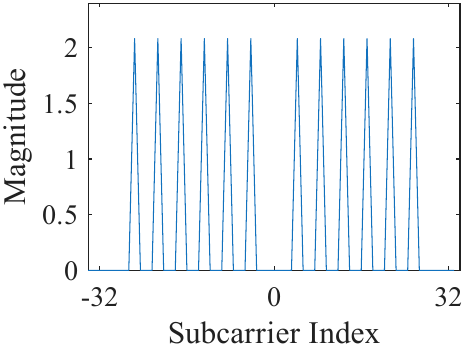}}
    \subfloat[]{\includegraphics[width=1.7in]{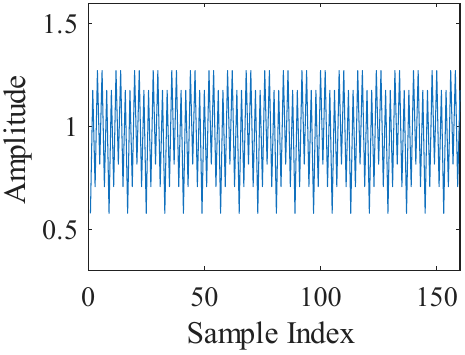}}    
	\caption{Visualization of the L-STF preamble. (a) Frequency-domain spectrum. (b) Time-domain waveform.}
	\label{fig:stf_visualization}
\end{figure}

To understand why L-STF can accommodate secret embedding without disrupting normal Wi-Fi operations, we examine its structure and functionality in detail. A visualization of the L-STF region is provided in Fig.~\ref{fig:stf_visualization}. 
In the frequency domain shown in Fig.~\ref{fig:stf_visualization}(a), a non-HT 20~MHz Wi-Fi packet consists of 64 subcarriers in total, among which the L-STF occupies 12 active subcarriers located at indices $\pm[24, 20, 16, 12, 8, 4]$. Applying a 64-point inverse fast Fourier transform (IFFT) to this frequency-domain structure generates four 16-sample time-domain STSs.
As shown in the time-domain waveform in Fig.~\ref{fig:stf_visualization}(b), the L-STF consists of ten identical STSs, each comprising 16 samples\footnote{The 64-point frequency-domain STS can be obtained by performing FFT on four consecutive time-domain STSs.}. Repeating one such STS ten times results in the 160-sample L-STF time-domain waveform shown in Fig.~\ref{fig:stf_visualization}(b).

The L-STF serves two critical functions by leveraging its periodic structure:
\begin{itemize}
    \item[(a)] \textbf{Packet Detection Functionality:} The receiver detects Wi-Fi signals through autocorrelation operations that identify the periodic characteristics of the L-STF. The delay autocorrelation of received repeated STSs is computed as:
    \begin{equation}\label{eq:packet_detection}
     P_n = \frac{\left|\sum_{k=0}^{15}y_{\mathrm{STS}}[n+k]y_{\mathrm{STS}}^*[n+k+16]\right|}{\sum_{k=0}^{15}|y_{\mathrm{STS}}[n+k+16]|^2},
    \end{equation}
    where $*$ denotes complex conjugation and $|\cdot|$ returns the magnitude. Given that the STS exhibits a repetition period of 16 samples, $P_n$ generates a plateau-like response upon L-STF detection.
    
    \item[(b)] \textbf{Coarse CFO Compensation Functionality:} The CFO is estimated by analyzing phase differences between adjacent periods. The CFO estimation is mathematically given as
    \begin{equation}\label{eq:cfo_compensation}
    \Delta \hat{f} = \frac{f_s}{2\pi \cdot 16}\angle\left(\sum_{k=0}^{N_{\mathrm{short}}-17}y_{\mathrm{STS}}[k]y_{\mathrm{STS}}^*[k+16] \right),
    \end{equation}
    where $\angle(\cdot)$ returns the phase, $N_{\mathrm{short}}=160$, and $f_s$ is the sample rate of 20~MHz. The estimated offset $\Delta \hat{f}$ is then used to compensate the received signal as
    \begin{equation}
    \label{Coarse_CFO_compensation}
    \tilde{y}[n] = y[n]\cdot e^{-j2\pi n \Delta \hat{f}/f_s},
    \end{equation}
    where $\tilde{y}[n]$ denotes the compensated signal.
\end{itemize}

The key insight is that both functionalities rely solely on the 16-sample periodicity of the L-STF. Therefore, by ensuring that the embedded secrets do not disrupt this periodicity, i.e., which is achieved by applying identical distortions to each repeated STS, the proposed scheme maintains full compatibility with standard Wi-Fi synchronization procedures. This design principle is validated through both simulation (Section~\ref{sec:impact_demodulation_sim}) and real-world experiments (Section~\ref{sec:impact_demodulation_exp}).

\subsection{Neural Network-based Secret Embedder}\label{sec:lstf_generation}

Building upon the understanding that L-STF periodicity must be preserved, this subsection describes how Alice embeds secret messages into the frequency-domain representation of an STS using an NN-based approach.

Alice embeds the secret message $\mathbf{s}$ into a single STS using an NN-based secret embedder. Specifically, the NN introduces imperceptible distortions to only the 12 active subcarriers of the frequency-domain STS representation $\mathbf{X}$, which is mathematically formulated as
\begin{equation}
    \tilde{\mathbf{X}}_{\mathrm{STS}} = \mathcal{N}_{\mathrm{emb}}(\mathbf{s}, \mathbf{X}_{\mathrm{STS}}; \boldsymbol{\theta}_{\mathrm{emb}}),
\end{equation}
where $\mathcal{N}_{\mathrm{emb}}(\cdot; \boldsymbol{\theta}_{\mathrm{emb}})$ represents the NN embedding function with learnable parameter set $\boldsymbol{\theta}_{\mathrm{emb}}$. The input $\mathbf{X}_{\mathrm{STS}}$ represents the 12 active subcarriers in the standard frequency-domain STS, given as
\begin{equation}
    \mathbf{X}_{\mathrm{STS}} = [ X^1_{\mathrm{STS}}, X^2_{\mathrm{STS}}, \dots, X^{12}_{\mathrm{STS}}], \quad \mathrm{where} \quad X^i_{\mathrm{STS}} \in \mathbb{C}.
\end{equation}
The remaining null subcarriers (i.e., DC null and guard bands) are zero-padded to ensure compatibility with existing Wi-Fi protocols. The output $\tilde{\mathbf{X}}_{\mathrm{STS}} \in \mathbb{C}^{12}$ denotes the frequency-domain STS after applying artificial distortion.

\begin{figure}[!t]
    \centering
    \includegraphics[width=3.4in]{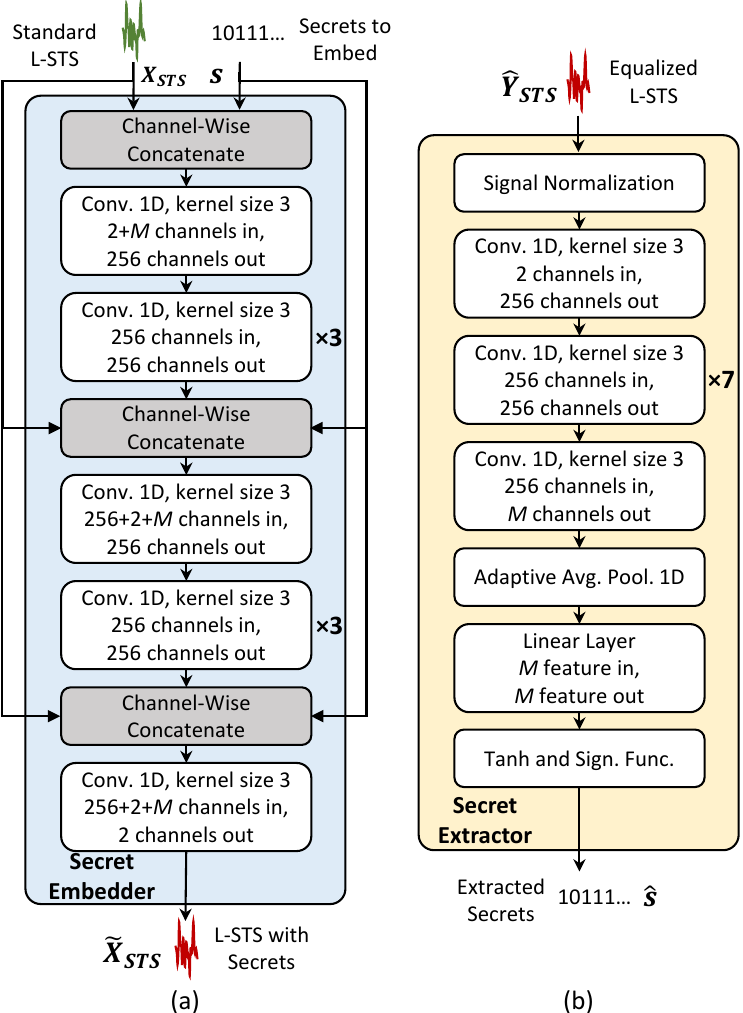}
    \caption{1D-CNN architectures used by the proposed system. (a) Secret embedder. (b) Secret extractor.}
    \label{fig:nn_architecture}
\end{figure}

\subsubsection{Network Architecture}\label{sec:network_architecture}
The secret embedder and the secret extractor are both implemented as one-dimensional convolutional neural networks (1D-CNNs) operating along the ordered positions of the 12 active L-STF subcarriers. The architecture of the secret embedder $\mathcal{N}_{\mathrm{emb}}(\cdot; \boldsymbol{\theta}_{\mathrm{emb}})$ is illustrated in Fig.~\ref{fig:nn_architecture}(a). Given the secret $\mathbf{s}$ and the standard frequency-domain STS $\mathbf{X}_{\mathrm{STS}}$ as input, the model produces the distorted STS $\tilde{\mathbf{X}}_{\mathrm{STS}}$ as output.

Following the channel-wise feature concatenation strategy of~\cite{zhu2018hidden}, we organize the network input into feature channels, where a channel here refers to a neural-network feature channel rather than a wireless propagation channel. Specifically, at each of the 12 active subcarrier positions, the real and imaginary parts of the complex STS coefficient constitute two signal-feature channels. The $M$-bit secret is repeated across all 12 positions and concatenated with these two channels, resulting in an input tensor of shape $(2+M,12)$. This formulation enables the network to learn a secret-bearing signal modification directly from data, instead of relying on a manually designed bit-to-signal mapping.

The input tensor is then processed by nine one-dimensional convolutional layers. The first eight form hidden convolutional blocks, each with 256 output channels, kernel size 3, same padding, and ReLU activation. The final convolutional layer also uses kernel size 3 and generates the two real and imaginary output channels of the distorted STS $\tilde{\mathbf{X}}_{\mathrm{STS}}$. In addition, the standard STS feature channels and the repeated secret are concatenated with the intermediate features before the second group of hidden blocks and again before the output layer. Through these skip connections, the unmodified subcarrier values and the secret remain accessible at every stage, allowing the network to determine the distortion magnitude with direct reference to the cover signal being perturbed.

\subsection{L-STF Generation}
\begin{figure}[!t]
	\centering
    \subfloat[]{\includegraphics[width=3.4in]{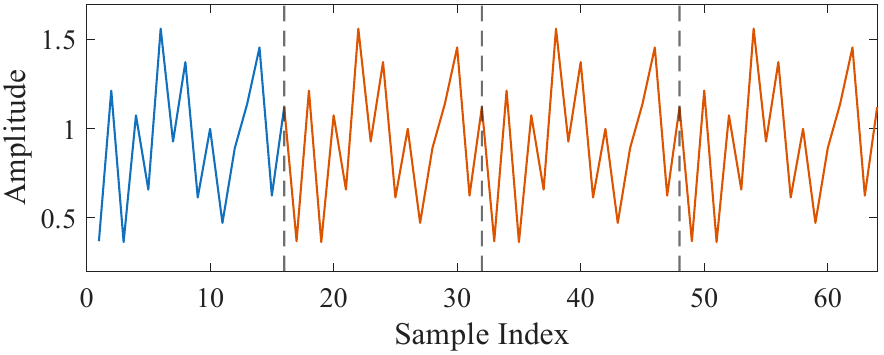}}\\[-1ex]
    \subfloat[]{\includegraphics[width=1.7in]{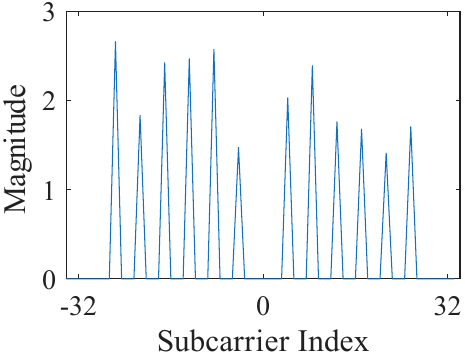}}
    \subfloat[]{\includegraphics[width=1.7in]{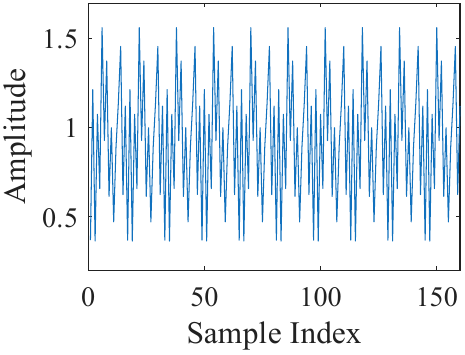}}    
	\caption{Visualization of the distorted L-STF preamble with a 14-bit secret embedded. (a) Distorted STS in the time domain (showing all four cyclic repetitions from IFFT output). (b) Distorted L-STF in the frequency domain. (c) Distorted L-STF in the time domain (showing all ten repetitions). } 
	\label{fig:distorted_stf_visualization}
\end{figure}

After obtaining the distorted frequency-domain STS, we convert it to the time domain and construct the complete L-STF through repetition. The 12 active frequency-domain subcarriers $\tilde{\mathbf{X}}_{\mathrm{STS}}$ are first zero-padded to 64 points and then converted into the time domain using an IFFT:
\begin{equation}
    \mathbf{\tilde{x}}_{\mathrm{STS}} = \mathcal{F}^{-1}\{\tilde{\mathbf{X}}_{\mathrm{STS}}\},
\end{equation}
where $\mathcal{F}^{-1}$ denotes the IFFT operation, and $\mathbf{\tilde{x}}_{\mathrm{STS}}$ represents the resulting distorted time-domain STS. It is important to note that the IFFT output contains four identical STSs, as illustrated in 
% It is important to note that the IFFT output contains four identical and cyclic STSs, \jz{what is cyclic STSs} as illustrated in 
Fig.~\ref{fig:distorted_stf_visualization}(a). We extract the first 16-sample STS for further processing.

To preserve the packet detection and CFO compensation functionalities discussed in Section~\ref{sec:function_lstf}, we repeat the distorted STS ten times to construct the complete L-STF:
\begin{equation}
    \mathbf{\tilde{x}}_{\mathrm{STF}} = \underbrace{[\mathbf{\tilde{x}}_{\mathrm{STS}}, \mathbf{\tilde{x}}_{\mathrm{STS}}, \ldots, \mathbf{\tilde{x}}_{\mathrm{STS}}]}_{10~ \mathrm{repetitions}},
\end{equation}
where $\mathbf{\tilde{x}}_{\mathrm{STF}}$ denotes the final L-STF containing the embedded secret. Crucially, because identical distortions are applied to all ten STSs, the 16-sample periodicity is perfectly preserved. Consequently, both the autocorrelation-based packet detection algorithm in~\eqref{eq:packet_detection} and the phase-difference-based coarse CFO compensation method in~\eqref{eq:cfo_compensation} remain fully effective.

% Finally, Alice concatenates the distorted L-STF $\mathbf{\tilde{x}}_{\mathrm{STF}}$ with the unmodified L-LTF, L-SIG, and DATA fields to form a protocol-compliant Wi-Fi packet, which is then transmitted over the air via a radio frequency front-end.

% \subsubsection{Visualization of the Distorted L-STF}

It is worth emphasizing that we embed secrets solely within the 12 active subcarriers, leaving the remaining 52 subcarriers (including DC null and guard bands) unmodified. This design is driven by two key considerations. First, introducing energy into inactive subcarriers would generate abnormal spectral patterns that deviate from standard Wi-Fi signal characteristics. Such anomalies are readily detectable by a warden through spectral analysis, thereby significantly elevating the risk of exposure. Second, since inactive subcarriers carry no legitimate signal energy in standard Wi-Fi transmissions, any covert information embedded therein must be transmitted at an extremely low power level to remain undetectable. Such low-power embedding would be overwhelmed by channel noise and interference, leading to an extremely poor SNR and thus severely compromising the reliability of secret recovery at the receiver.

To demonstrate the imperceptibility of the secret embedding, Fig.~\ref{fig:distorted_stf_visualization} provides a visualization of the distorted L-STF waveform with a 14-bit secret embedded.
Comparing with the standard L-STF shown in Fig.~\ref{fig:stf_visualization}, we observe varying degrees of amplitude modulation across the 12 active subcarriers in Fig.~\ref{fig:distorted_stf_visualization}(b), which encode the embedded secret information.
The time-domain waveform exhibits only slight distortion, as illustrated in Fig.~\ref{fig:distorted_stf_visualization}(c). The overall periodic structure remains clearly visible, confirming that the 16-sample repetition pattern is preserved.

\subsection{Wi-Fi Modulation and Packet Construction}
The remaining fields of the Wi-Fi packet are generated strictly adhering to the IEEE 802.11 standard to ensure normal communication. Specifically, the L-LTF is generated using the standard sequences to facilitate channel estimation at the receiver. The L-SIG is constructed to encode the rate and length parameters derived from the payload. The DATA field is produced by passing the payload bits through the standard transmit chain, including scrambling, convolutional encoding, interleaving, and modulation.

Finally, the distorted L-STF $\mathbf{\tilde{x}}_{\mathrm{STF}}$ is concatenated with these standard waveform components in the time domain. The complete packet waveform is then up-converted and transmitted via the RF front-end. This assembly process guarantees that the non-steganographic regions of the packet remain pristine, thereby supporting the standard demodulation procedures described in Section~\ref{sec:bob}.

\section{Secrets Extraction at Bob}\label{sec:bob}
This section details the secret extraction workflow at the receiver. After standard synchronization and payload demodulation, Bob performs L-STF equalization to mitigate multipath-induced frequency-selective fading. The equalized PHY Wi-Fi signal is then fed into an NN-based secret extractor to recover secret messages.

\subsection{Wi-Fi Demodulation}

The receiver Bob captures the packet waveform transmitted by Alice and performs standard Wi-Fi demodulation procedures.
First, Bob executes packet detection and coarse CFO compensation using the received L-STF, as described in Section~\ref{sec:function_lstf}. As discussed earlier, these two functionalities remain fully operational despite the secret embedding, since the 16-sample periodic repetition property is preserved through identical distortions applied to all ten STSs.

Following coarse synchronization, Bob utilizes the L-LTF for fine time and frequency synchronization as well as \textbf{channel estimation}, which is an integral step of standard Wi-Fi reception. In particular, the channel state information (CSI) is estimated from the L-LTF using the least-squares (LS) algorithm:
\begin{equation}
    \hat{\mathbf{H}} = \frac{\mathbf{Y}_{\mathrm{LTS}}}{\mathbf{X}_{\mathrm{LTS}}},
\end{equation}
where $\hat{\mathbf{H}} \in \mathbb{C}^{52}$ represents the estimated CSI across all data subcarriers, $\mathbf{Y}_{\mathrm{LTS}}$ denotes the received frequency-domain L-LTS, and $\mathbf{X}_{\mathrm{LTS}}$ represents the standard frequency-domain L-LTS known to the receiver.

The estimated CSI $\hat{\mathbf{H}}$ is then applied to equalize the L-SIG and DATA fields, enabling reliable payload recovery. Because the proposed PHY distortion is confined exclusively to the L-STF region, while the L-LTF, L-SIG, and DATA fields remain unmodified, the proposed secret embedding scheme does not impair the primary Wi-Fi link. In other words, standard Wi-Fi demodulation performance is preserved.

\subsection{L-STF Equalization}

After successfully recovering the payload, Bob turns to extracting the embedded secret from the secret-bearing L-STF. 
Since Wi-Fi is a wideband system, the received signal typically experiences multipath-induced frequency-selective fading, which can distort the secret-bearing subcarriers. To ensure reliable secret recovery under diverse channel conditions, we \textbf{reuse the CSI $\hat{\mathbf{H}}$ estimated from the L-LTF during Wi-Fi demodulation} and extend channel equalization to the L-STF region, referred to as \textbf{L-STF equalization}.

Specifically, for each received STS, Bob converts the time-domain segment $\mathbf{y}_{\mathrm{STS}}$ into the frequency domain via FFT to obtain $\mathbf{Y}_{\mathrm{STS}}$, extracts the 12 active subcarriers corresponding to L-STF, and performs equalization as
\begin{equation}
    \hat{\mathbf{Y}}_{\mathrm{STS}} = \frac{\mathbf{Y}_{\mathrm{STS}}}{\hat{\mathbf{H}}_{\mathrm{STS}}},
\end{equation}
where $\hat{\mathbf{Y}}_{\mathrm{STS}} \in \mathbb{C}^{12}$ denotes the channel-equalized STS, and $\hat{\mathbf{H}}_{\mathrm{STS}}$ denotes the CSI values at the 12 L-STF subcarrier indices.

By applying equalization to the L-STF, we effectively remove channel-induced amplitude and phase distortions, leaving primarily the secret-bearing artificial distortions introduced by Alice. This significantly enhances the robustness of secret extraction, particularly in multipath-rich environments.

\subsection{Neural Network-Based Secret Extraction}\label{sec:secret_extraction_nn}

After L-STF equalization, the equalized STS is fed into a pre-trained NN, termed the secret extractor, to recover the embedded secrets. The extraction process is mathematically expressed as
\begin{equation}
    \hat{\mathbf{s}} = \mathcal{N}_{\mathrm{ext}}(\hat{\mathbf{Y}}_{\mathrm{STS}};\boldsymbol{\theta}_{\mathrm{ext}}),
\end{equation}
where $\hat{\mathbf{s}} \in \{-1,+1\}^M$ represents the extracted $M$-bit secret, and $\mathcal{N}_{\mathrm{ext}}(\cdot;\boldsymbol{\theta}_{\mathrm{ext}})$ denotes the NN operation with parameter set $\boldsymbol{\theta}_{\mathrm{ext}}$.

The secret extractor is also built upon a 1D-CNN, as illustrated in Fig.~\ref{fig:nn_architecture}(b). To construct its input, the equalized frequency-domain STS is first normalized, and its real and imaginary parts are then organized into an input tensor of shape $(2,12)$. The tensor is processed by nine one-dimensional convolutional layers. The first eight layers each use 256 output channels, kernel size 3, same padding, and ReLU activation, while the ninth layer uses kernel size 3 and maps the features to $M$ channels. Subsequently, an adaptive average pooling layer aggregates the features over the 12 subcarrier positions into an $M$-element vector, which is then transformed by an $M$-input/$M$-output fully connected layer into the final $M$-element output feature vector.

Finally, a Tanh activation function maps the output features to soft symbol estimates within the range $(-1,+1)$, and a sign function converts them into the hard-decision binary sequence $\hat{\mathbf{s}}$. The Tanh activation provides smooth gradients that facilitate training, while the sign function yields robust bit decisions during inference.

\section{Learning to Approach Optimal Steganographic Capacity}\label{sec:training}

The secret embedder and extractor described in Sections~\ref{sec:alice} and~\ref{sec:bob} must be carefully designed to achieve high steganographic capacity while maintaining covertness and robustness. Rather than relying on manually crafted embedding rules, we adopt a data-driven approach where both NNs are jointly trained in an end-to-end manner on a trustworthy simulation platform. Upon completion of training, the learned models are securely distributed to Alice and Bob for practical deployment. This section describes the training methodology and the loss functions that guide the learning process.

\subsection{End-to-End Training with Wireless Channel Simulator}

\begin{figure}[!t]
    \centering
    \includegraphics[width=2.2in]{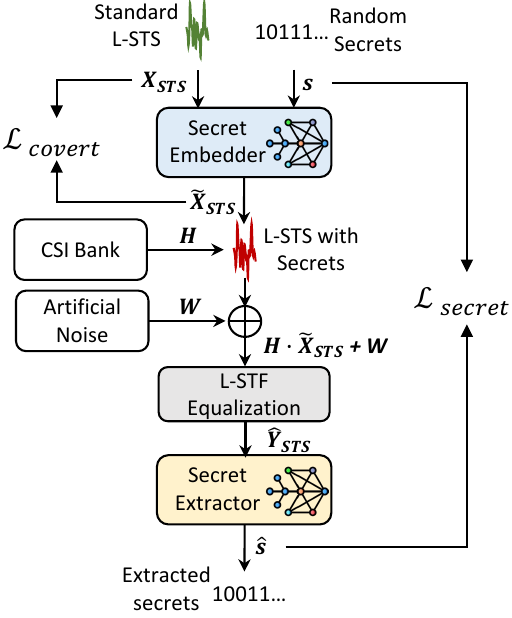}
    \caption{End-to-end training architecture for the secret embedder and extractor. The training incorporates a TGn channel simulator and L-STF equalization to enhance robustness against real-world channel variations.}
    \label{fig:nn_training}
\end{figure}

The training process is conducted in an end-to-end manner, as illustrated in Fig.~\ref{fig:nn_training}. The secret embedder and extractor are concatenated to form an encoder-decoder architecture, enabling the emulation of a complete secret transmission and recovery process. To enhance the robustness to channel variations, we specifically integrate the IEEE 802.11 TGn channel model as a data augmentation technique to simulate various multipath effects during training. 

The training procedure consists of the following steps: 
% \jz{channel simulation is not shown in Fig. 6. other blocks do not have the same name.} \blue{change Wireless Channel Simulator to CSI Bank, others remain unchanged }
\begin{enumerate}
    \item[(a)] \textbf{Secret Embedder:} A random $M$-bit binary sequence $\mathbf{s}$ is generated and embedded into the standard frequency-domain STS $\mathbf{X}_{\mathrm{STS}}$ using the secret embedder, producing the distorted STS $\tilde{\mathbf{X}}_{\mathrm{STS}}$.
    
    \item[(b)] \textbf{CSI Bank and Artificial Noise:} To simulate realistic wireless propagation, a CSI $\mathbf{H}$ is randomly selected from a pre-constructed CSI bank and element-wise multiplied with the distorted STS. AWGN $\mathbf{W}$ is then added to model receiver noise:
    \begin{equation}
        \mathbf{Y}_{\mathrm{STS}} = \mathbf{H} \odot \tilde{\mathbf{X}}_{\mathrm{STS}} + \mathbf{W},
    \end{equation}
    where $\odot$ denotes element-wise multiplication, and the SNR is uniformly sampled from the range $[10,30]$~dB to cover typical operating conditions. 
    
    The CSI bank is constructed offline using the IEEE 802.11 TGn channel Model-B provided in the MATLAB WLAN Toolbox~\cite{wlanTGnChannel}, which models typical residential indoor environments with moderate multipath characteristics. We simulated 1,000 independent channel realizations and recorded the corresponding frequency-domain CSI measurements across the 12 L-STF subcarriers to create this bank. During training, CSI samples are randomly drawn from this bank to apply diverse channel conditions on the PHY signal.
    
    \item[(c)] \textbf{L-STF Equalization:} The received signal $\mathbf{Y}_{\mathrm{STS}}$ undergoes L-STF equalization to produce the equalized signal $\hat{\mathbf{Y}}_{\mathrm{STS}} = \mathbf{Y}_{\mathrm{STS}} / \mathbf{H}$. This step mimics the equalization process that will be performed at the real receiver.
    
    \item[(d)] \textbf{Secret Extractor:} The equalized signal $\hat{\mathbf{Y}}_{\mathrm{STS}}$ is fed into the secret extractor to produce the recovered secret $\hat{\mathbf{s}}$.
    
    % \item[(e)] \textbf{Loss Computation:} The total loss $\mathcal{L}$ is calculated based on the discrepancy between the transmitted secret $\mathbf{s}$ and the recovered secret $\hat{\mathbf{s}}$, as well as the distortion level between the original and distorted STS. The gradients are then backpropagated through both networks to update their parameters.
\end{enumerate}

This end-to-end training paradigm enables the embedder and extractor to co-adapt, allowing the embedder to learn distortion patterns that the extractor can reliably decode even after channel corruption.

\subsection{Loss Functions}

% We employ a data-driven approach to learn the secret embedding rule that achieves the optimal steganographic capacity. 

Instead of relying on fixed, hand-crafted embedding rules that may be suboptimal, we allow the NNs to autonomously discover the most effective embedding strategy by minimizing a carefully designed composite loss function. This loss function balances two objectives: maximizing secret recovery accuracy and minimizing the perceptibility of waveform distortions.

\subsubsection{Secret Recovery Loss}
The secret loss $\mathcal{L}_{\mathrm{secret}}$ ensures accurate secret recovery by measuring the discrepancy between the transmitted secret $\mathbf{s}$ and the recovered secret $\hat{\mathbf{s}}$:
\begin{equation}
    \mathcal{L}_{\mathrm{secret}} = \frac{\|\hat{\mathbf{s}} - \mathbf{s}\|_2}{M}, 
\end{equation}
where $\|\cdot\|_2$ denotes the L2 norm, and $M$ is the number of secret bits. This loss directly penalizes bit errors and drives the networks to minimize the secret bit error rate (SBER).

\subsubsection{Covertness Loss}
The covert loss $\mathcal{L}_{\mathrm{covert}}$ limits the magnitude of distortions by enforcing high similarity between the distorted STS $\tilde{\mathbf{X}}_{\mathrm{STS}}$ and the standard STS $\mathbf{X}_{\mathrm{STS}}$:
\begin{equation}
    \mathcal{L}_{\mathrm{covert}} = \frac{\|\tilde{\mathbf{X}}_{\mathrm{STS}} - \mathbf{X}_{\mathrm{STS}}\|_2}{K},
\end{equation}
where $K=12$ represents the number of active subcarriers in the L-STF. This loss prevents the embedder from introducing excessive distortions that might be detectable by a warden or degrade Wi-Fi synchronization performance.

\subsubsection{Composite Loss}
The total loss function $\mathcal{L}$ is a weighted combination of the secret loss and covert loss:
\begin{equation}
    \mathcal{L} = \gamma \cdot \mathcal{L}_{\mathrm{covert}} + (1-\gamma) \cdot \mathcal{L}_{\mathrm{secret}},
\end{equation}
where $\gamma \in [0,1]$ is a predefined weighting parameter that controls the trade-off between covertness and secret recovery performance. A higher value of $\gamma$ emphasizes imperceptibility at the cost of reduced capacity, while a lower value prioritizes capacity at the risk of increased detectability. In this work, we empirically set $\gamma = 0.2$.

\section{Simulation Evaluation}\label{sec:simulation_results}

This section conducts comprehensive simulations to preliminarily validate the proposed PHY information hiding scheme against the three core challenges discussed in Section~\ref{sec:challenges}.

\subsection{Simulation Setting}\label{sec:nn_training_configuration}

\subsubsection{Wi-Fi PHY}

% The Wi-Fi PHY is implemented with the MATLAB WLAN toolbox~\cite{wlanToolbox}. IEEE 802.11 20~MHz non-HT Wi-Fi packets are generated and transmitted over-the-air using GNU Radio software on a USRP B210 platform. Specifically, we generate Wi-Fi beacon frames and embed random $M$-bit secrets within the L-STF portion, where the secret length $M$ is specified in the following subsections. For evaluation purposes, the embedded secrets are also included in the service set identifier (SSID) field as plaintext, serving as ground truth for calculating evaluation metrics such as the secret bit error rate. 
The Wi-Fi PHY layer is implemented using the MATLAB WLAN Toolbox~\cite{wlanToolbox}. At the transmitter, standard 20~MHz SISO non-HT waveforms are generated. Simultaneously, $M$-bit random secret sequences are embedded into the L-STF region following the embedding process defined in Section~\ref{sec:alice}. The generated Wi-Fi baseband signals are then passed through a TGn channel model to simulate wireless propagation effects. At the receiver, the signals are processed through standard Wi-Fi reception stages, including packet detection, synchronization, and demodulation, etc. Subsequently, the L-STF portion is extracted and fed into the secret extractor for secret recovery.

% \blue{The L-SIG and DATA fields are dynamically synthesized using MATLAB to encapsulate the variable SSID payload. In contrast, the L-STF and L-LTF are initialized as static sequences strictly adhering to the IEEE 802.11 standard.}
% A total of 10,000 Wi-Fi packets are transmitted for each experiment unless otherwise specified.

\subsubsection{Neural Network Training Configuration}

The NNs are implemented in the PyTorch library and trained using the RMSprop optimizer with a batch size of 64. The initial learning rate is set to $1\times 10^{-4}$, and a reduce-on-plateau scheduler is employed to adaptively adjust the learning rate by multiplying it by 0.1 when the validation loss fails to decrease for 30 consecutive epochs. To prevent overfitting, early stopping is utilized and triggered after 50 epochs without validation loss improvement. The hyperparameter $\gamma$ is set to 0.2 in this work. The NNs were trained on a workstation equipped with an Intel Core i5-12600KF CPU, 16 GB of RAM, and an NVIDIA GeForce RTX 3070 GPU.

\subsubsection{Evaluation Metric}

% We defined two metrics to evaluate the system performance, namely BER and PER.

% \subsubsection{Secret BER}
The SBER is utilized for evaluation, which measures the bit-level differences between the embedded and extracted secret messages, given as
\begin{equation}
    SBER = \frac{\mathcal{D}(\mathbf{s},\hat{\mathbf{s}})}{M},
\end{equation}
where $\mathcal{D}(\cdot, \cdot)$ computes the Hamming distance. The SBER directly evaluates the bit-level system performance.

Every simulation condition is evaluated with ten independently trained embedder--extractor pairs that differ only in their random seed, so that the reported performance is not tied to a single training run. Error bars are shown in the figures.

% The PER serves as a key metric for evaluating the reliability of the covert communication channel. It is defined as the ratio of incorrectly received packets to the total number of received packets, given by
% \begin{equation}
%     PER = \frac{N_{\mathrm{err}}}{N_{\mathrm{total}}},
% \end{equation}
% where $N_{\mathrm{err}}$ denotes the number of packets in which the embedded secret cannot be correctly recovered, i.e., $\mathbf{s} \neq \hat{\mathbf{s}}$, and $N_{\mathrm{total}}$ is the total number of received packets. A lower PER indicates a more reliable covert channel.

% \subsubsection{Compatibility}
% \red{A key requirement of our proposed method is compatibility with existing Wi-Fi systems, achieved by preserving the original functionality of the L-STF. To evaluate this, we conduct a toy simulation to compare its performance in packet detection and coarse CFO estimation against the standard L-STF. In addition, we demonstrate its practical applicability using a commodity SDR platform.}
% \blue{I suggest deleting this paragraph. An evaluation metric must be quantifiable, such as accuracy, precision, mean square error, etc. Our evaluation on compatibility is not quantifiable}
% \blue{To assess the feasibility of the proposed Wi-Fi information hiding scheme, we conducted a series of simulation-based pilot studies. These studies address two critical questions: (1) Does the introduced distortion impair the primary Wi-Fi link? (2) Can the receiver reliably extract embedded secrets after propagation through the wireless channel? The evaluation methodology is detailed below.}

\subsection{Validation of Normal Wi-Fi Demodulation and Reception}\label{sec:impact_demodulation_sim}

\begin{figure}[!t]
    \centering
    \subfloat[]{\includegraphics[width=3.4in]{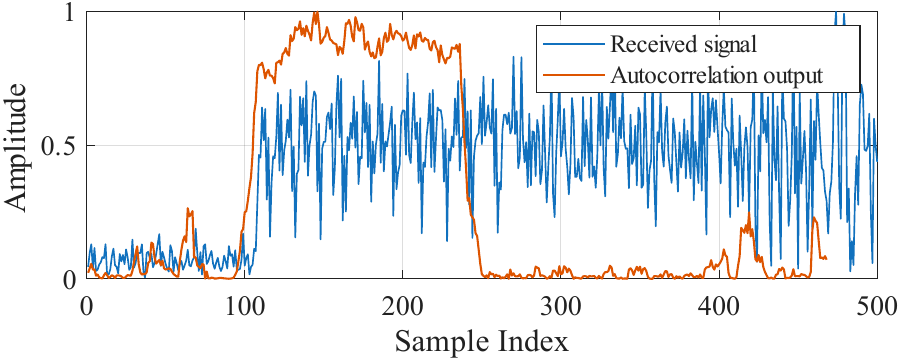}}\\
    \subfloat[]{\includegraphics[width=3.4in]{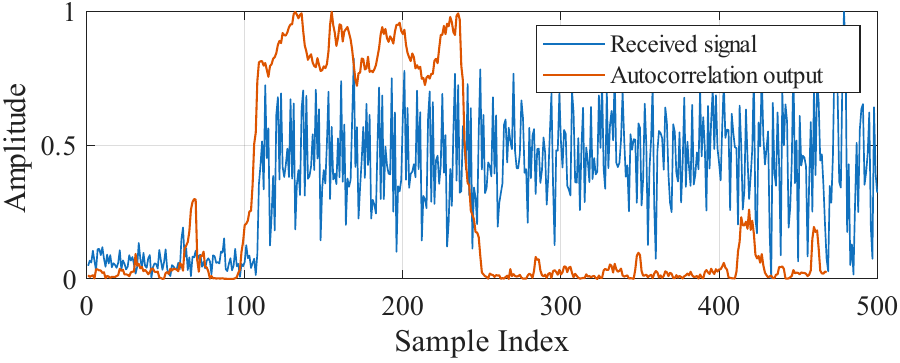}}
    \caption{Wi-Fi packet detection results at 15~dB. (a) Autocorrelation output of standard L-STF. (b) Autocorrelation output of distorted L-STF.}
    \label{fig:packet_detection}
\end{figure}

A fundamental requirement of any Wi-Fi PHY-layer information hiding scheme is that the covert channel must not interfere with the primary communication link, such that the transmitted signals remain indistinguishable from standard Wi-Fi waveforms to legacy receivers and potential wardens. In the proposed design, as described in Section~\ref{sec:alice}, the distorted STS is repeated ten times to form the L-STF preamble. This construction strictly preserves the periodic repetition structure of the standard L-STF, which serve as the foundation for its essential synchronization functionalities, including packet detection and coarse CFO compensation, as reviewed in Section~\ref{sec:bob}. Since the payload waveform is left entirely unmodified, the demodulation and decoding of Wi-Fi data symbols are inherently unaffected.

To verify the overall integrity of normal Wi-Fi reception, we conducted extensive simulations using the MATLAB WLAN Toolbox. Specifically, 10,000 Wi-Fi packets were transmitted over the TGn channel Model-B with SNR values ranging from 10 to 25~dB, while random secrets of varying lengths were embedded into the L-STF preamble. Across all simulation trials, no packet loss was observed. \textbf{All transmitted packets were successfully detected, and the frame check sequence (FCS) validation of the payload consistently passed}, providing strong evidence that the proposed information hiding scheme does not compromise standard Wi-Fi demodulation and decoding.
Building upon this overall validation, we next examine in detail whether the two core functionalities of the L-STF, i.e., packet detection and coarse CFO compensation, are individually preserved.

\subsubsection{Impact on Packet Detection Functionality}

We first investigate the impact of the proposed PHY information hiding scheme on correlation-based packet detection. Fig.~\ref{fig:packet_detection} compares the autocorrelation outputs of the distorted L-STF with those of the standard L-STF. Results from 10,000 independent simulation trials show that the injected PHY-level distortions strictly preserve the inherent periodicity and correlation characteristics of the L-STF. Consequently, the distinctive plateau structure required for reliable packet detection and coarse timing synchronization remains unaffected in all cases. These observations confirm that embedding secret information into the L-STF does not degrade the packet detection performance of legacy Wi-Fi receivers.

\begin{figure}[!t]
    \centering
    \includegraphics[width=3.4in]{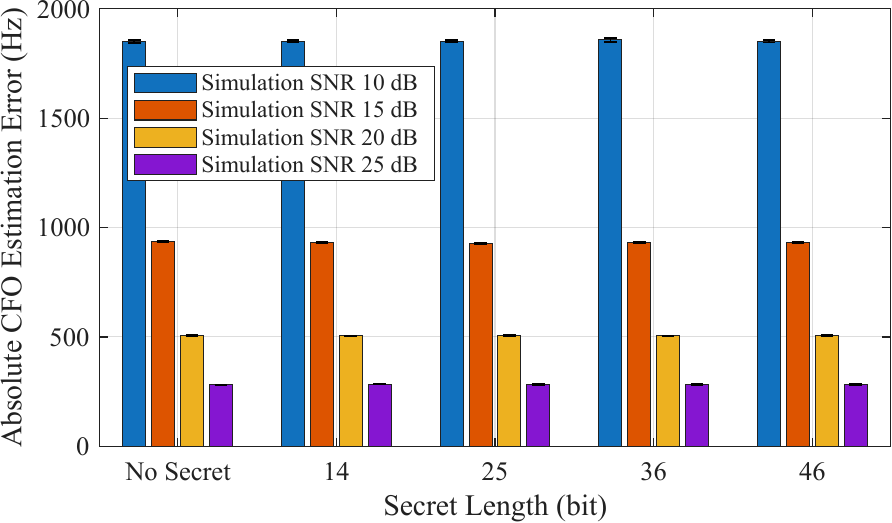}
    \caption{Simulation results of CFO estimation.}
    \label{fig:impact_CFO}
\end{figure}

\subsubsection{Impact on Coarse CFO Compensation Functionality}

We further evaluate the effect of the proposed scheme on coarse CFO estimation.
To this end, random CFO offsets were introduced into each of the 10,000 transmitted Wi-Fi packets, with the CFO uniformly distributed within $\pm2$~ppm, corresponding to approximately $\pm5,000$~Hz at the 2.4~GHz carrier frequency.
At the receiver, the CFO was estimated using the L-STF, and the resulting mean absolute estimation errors are presented in Fig.~\ref{fig:impact_CFO}.
The results indicate that the CFO estimation error is independent of the secret length and is instead primarily governed by the SNR.
This confirms that the coarse CFO compensation functionality of the L-STF remains preserved under the proposed PHY information hiding scheme.

\subsection{Evaluation of Steganographic Capacity}
\begin{figure}[!t]
    \centering
    \includegraphics[width=3.4in]{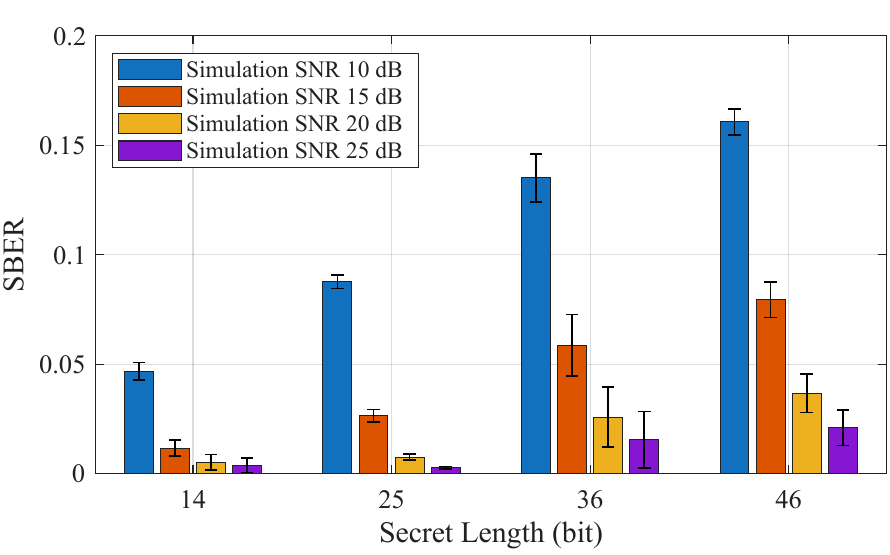}
    \caption{Simulation result of steganographic capacity.}
    \label{fig:simluation_ber}
\end{figure}

Building upon the previous subsection's validation that normal Wi-Fi demodulation remains unaffected by PHY-layer information hiding, this section evaluates the steganographic capacity of the proposed scheme.
The simulation is conducted using IEEE 802.11 TGn channel Model-B at SNR levels from 10 to 25~dB, with secret lengths ranging from 14 to 46 bits.
As shown in Fig.~\ref{fig:simluation_ber}, the SBER remains below 0.021 at 25~dB SNR.
More importantly, the results reveal a clear trade-off between capacity and reliability: SBER grows with the embedded secret length, since a longer secret must be carried by the same 12 subcarriers under the same distortion budget.
The SBER reported here is measured before any error correction, so the residual errors at the longer secret lengths represent the working point on which an outer FEC code would operate.
These simulations establish the feasibility of the proposed scheme and provide a basis for the subsequent hardware prototype experiments.

\subsection{Evaluation of Robustness to Channel Variations}

% \begin{figure}[!t]
%     \centering
%     \includegraphics[width=3.4in]{figures/simulation_ber_stitched_v2.pdf}
%     \caption{Simulation results under various channel effects, i.e., TGn channel models B to F are utilized. \jz{should be L-STF. might be better to use a table.} \blue{See TABLE I and II}}
%     \label{fig:simulation_channel_robust_ber}
% \end{figure}

% \begin{table*}[!t]
%     \centering
%     \caption{Comparison of SBER Across Different Channel Models}
%     \label{tab:sber_performance}
%     \begin{tabular}{lccccc}
%         \toprule
%         \multirow{2}{*}{\textbf{Equalization Scheme}} & \multicolumn{5}{c}{\textbf{TGn Channel Model}} \\
%          & \textbf{Model-B} & \textbf{Model-C} & \textbf{Model-D} & \textbf{Model-E} & \textbf{Model-F} \\
%         \midrule
%         With L-STF Equalization    & $1.2 \times 10^{-4}$ & $2.5 \times 10^{-4}$ & $5.1 \times 10^{-3}$ & $8.4 \times 10^{-3}$ & $1.2 \times 10^{-2}$ \\
%         Without L-STF Equalization & $3.5 \times 10^{-2}$ & $8.9 \times 10^{-2}$ & $1.5 \times 10^{-1}$ & $2.2 \times 10^{-1}$ & $3.4 \times 10^{-1}$ \\
%         \bottomrule
%     \end{tabular}
% \end{table*}

\begin{figure}[!t]
    \centering
    \includegraphics[width=3.4in]{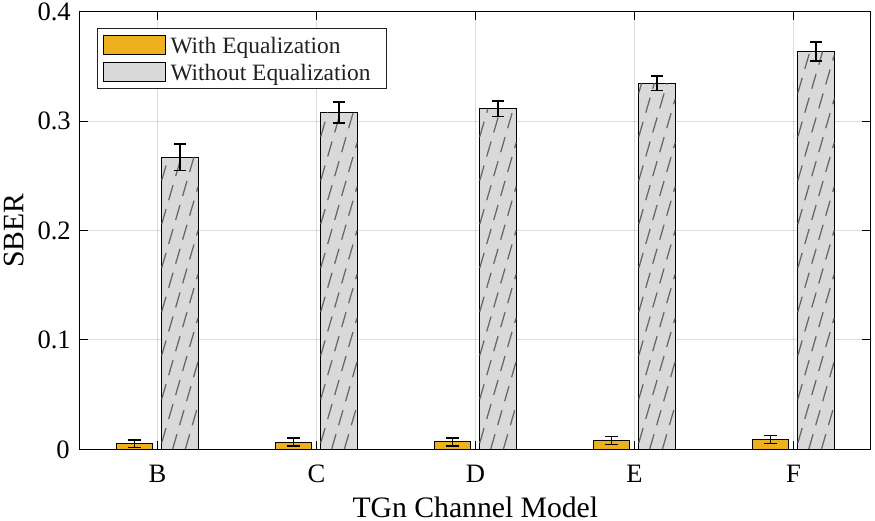}
    \caption{Effect of L-STF equalization on SBER under TGn channel models B--F at 20~dB received-waveform SNR for 14-bit secrets.}
    \label{fig:simulation_channel_robust_ber}
\end{figure}

% To evaluate the robustness of the proposed system against multipath propagation \blue{with the help of L-LTF equalization}, we conducted evaluations across a range of standardized channel models. \blue{The secret length and SNR were fixed at 14 bits and 20~dB, respectively. We utilized TGn channel models B through F, which model typical indoor environments with progressively more severe delay spreads and multipath fading characteristics.
% The simulation results are presented in Fig. \ref{fig:simulation_channel_robust_ber}, demonstrating that the system maintains consistent secret recovery capability across all tested channel conditions. Specifically, even under the challenging multipath profiles of models E and F, the SBER for a 14-bit secret remains below 0.004. These results indicate that the joint use of the neural network-based embedding strategy and L-LTF equalization effectively mitigates channel-induced distortions across diverse propagation environments.}
To evaluate the robustness of the proposed system against multipath propagation, we tested it under standardized TGn channel models B--F, which represent indoor environments with progressively increasing delay spread and multipath severity. The secret length and SNR were fixed at 14 bits and 20~dB, respectively.

As shown in Fig.~\ref{fig:simulation_channel_robust_ber}, with L-STF equalization the SBER remains below 0.009 across all evaluated TGn channel models, degrading only gradually from Model-B to the most dispersive Model-F.
Without L-STF equalization the SBER exceeds 0.26 under every channel model, which leaves the secret unrecoverable.
These results demonstrate the importance of L-STF equalization for mitigating multipath-induced distortion.

\subsection{Evaluation of Robustness under Attacks}\label{sec:attack_robustness}
To evaluate the robustness of the proposed PHY information hiding scheme under attacks, we consider two cases: a passive eavesdropper who attempts to read the secret, and an active jammer who attempts to prevent recovery.
Both evaluations use IEEE 802.11 TGn channel Model-B at 20~dB SNR, with secret lengths ranging from 14 to 46 bits.

\subsubsection{Eavesdropping}\label{sec:eavesdropping}

\begin{figure}[!t]
    \centering
    \includegraphics[width=3.4in]{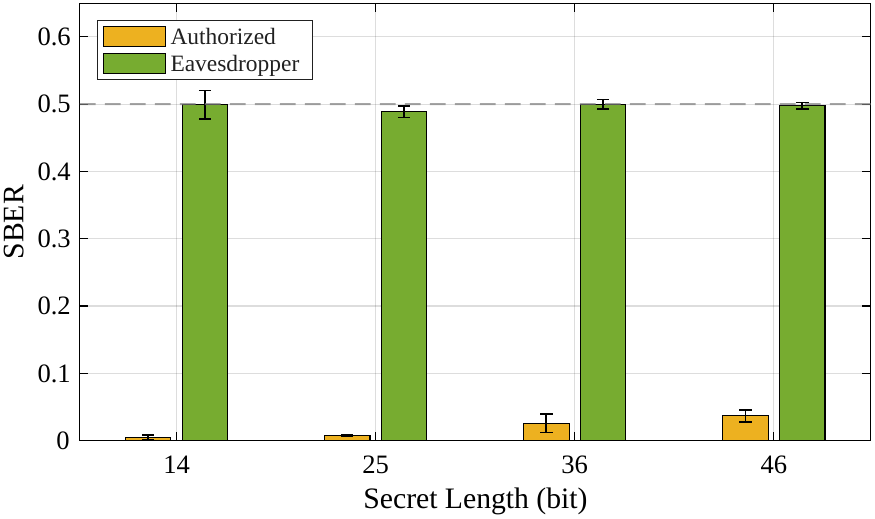}
    \caption{Eavesdropping evaluation under TGn Model-B at 20~dB SNR.}
    \label{fig:sim_eavesdrop}
\end{figure}

We consider a passive eavesdropper, Eve, who knows the proposed method, the network architecture, and the entire training procedure, but does not hold the weights shared by Alice and Bob.
Eve can only train an embedder--extractor pair independently from scratch.
As shown in Fig.~\ref{fig:sim_eavesdrop}, this mismatched extractor yields an SBER near $0.5$ at every secret length, which is indistinguishable from random guessing.
In contrast, the authorized extractor achieves SBERs of 0.0052, 0.0075, 0.0257, and 0.0366 for 14-, 25-, 36-, and 46-bit secrets, respectively.
Knowledge of the design is therefore not sufficient to read the secret, and the trained weights act as the shared key.

\subsubsection{Jamming}\label{sec:jamming}

\begin{figure}[!t]
    \centering
    \includegraphics[width=3.4in]{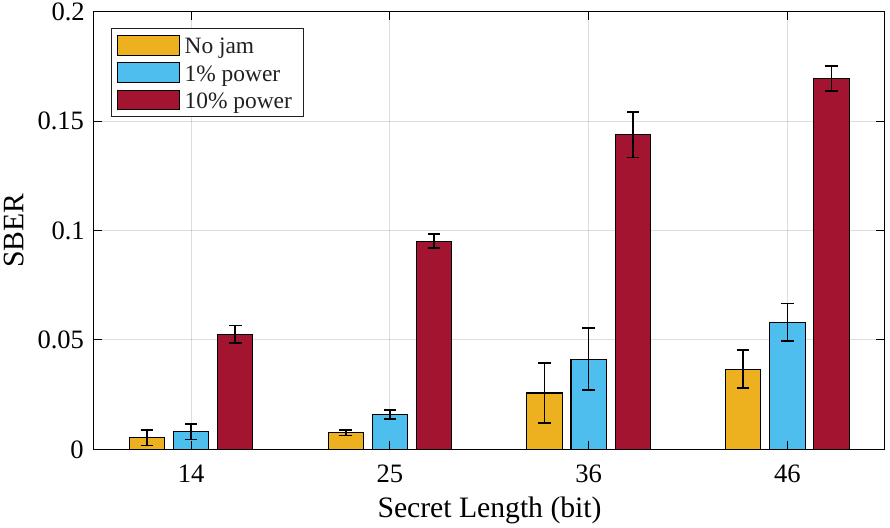}
    \caption{Additive jamming evaluation under TGn Model-B at 20~dB SNR.}
    \label{fig:sim_jamming}
\end{figure}

We next consider a jammer that continuously injects Gaussian noise to interfere with the communication link.
Fig.~\ref{fig:sim_jamming} compares the no-interference baselines with interference powers of 1\% and 10\% of the received signal power.
These two levels correspond to jamming-to-signal ratios of $-20$~dB and $-10$~dB, respectively.
Relative to the baselines of 0.0052, 0.0075, 0.0257, and 0.0366, the SBER increases to 0.0080, 0.0159, 0.0412, and 0.0580 at $-20$~dB, and to 0.0526, 0.0951, 0.1438, and 0.1694 at $-10$~dB.
The SBER increases gradually with the jamming power.
At $-20$~dB, the covert channel remains close to the baseline at every secret length.
At $-10$~dB, the shorter secrets remain usable, whereas the 46-bit case reaches an SBER of 0.1694, at which the secret can no longer be read directly and an outer FEC code becomes necessary.
Such a jamming attack, however, also degrades the primary Wi-Fi link.

\section{Experimental Evaluation}\label{sec:experimental_results}

This section further validates the proposed PHY information hiding scheme by implementing a hardware prototype and conducting over-the-air experiments.

\subsection{Experimental Setting}

We established an over-the-air testbed using two USRP B210 SDR platforms controlled by GNU Radio software. At the transmitter, IEEE 802.11 20 MHz non-HT beacon frames are generated with the MATLAB WLAN Toolbox, into whose L-STF region a random $M$-bit secret is embedded. To facilitate performance evaluation, the same secret sequence is encoded in plaintext within the SSID field, serving as the ground truth for SBER calculation. At the receiver, raw I/Q samples are continuously captured and processed offline to perform synchronization and extract the received L-STF segment. This received L-STF is then fed into the secret extractor for secret recovery. 
The hardware prototype uses the same NN architecture as the simulation study. For each secret length, one simulation-trained model is deployed without over-the-air fine-tuning.

\subsection{Validation of Normal Wi-Fi Demodulation and Reception}\label{sec:impact_demodulation_exp}
\begin{figure}[!t]
    \centering
    \includegraphics[width=3.4in]{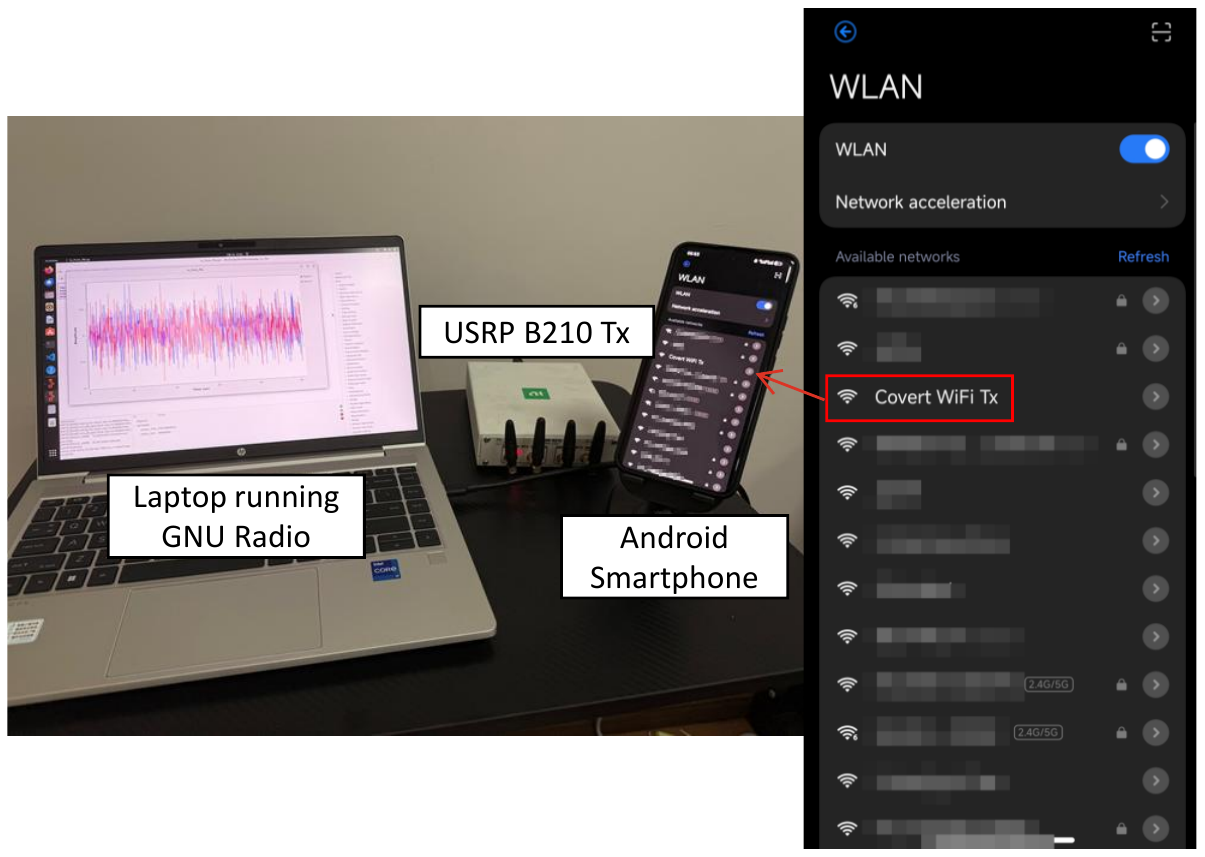}
    \caption{Evaluation of the impact on the primary Wi-Fi link.
    The USRP B210 continuously transmits Wi-Fi beacon frames with a random 14-bit secret embedded. These frames are successfully received and decoded by a commodity Android smartphone.}
    \label{fig:phone_receiving}
\end{figure}

We further validate that the primary Wi-Fi channel is unaffected using the hardware prototype. As shown in Fig.~\ref{fig:phone_receiving}, the USRP B210 is continuously transmitting Wi-Fi beacon frames, and a random 14-bit secret is embedded within the L-STF preamble. A commodity Android smartphone is then used to receive these beacon frames. As shown on the smartphone screen, the beacon frames are successfully detected and demodulated. The USRP B210 transmitter is correctly recognized in the Wi-Fi discovery interface with the SSID labeled as `Covert Wi-Fi Tx'. The compatibility with commercial smartphones provides additional evidence that the proposed PHY information hiding scheme does not interfere with the normal operation of the primary Wi-Fi link. 

\subsection{Evaluation of Steganographic Capacity}\label{sec:capacity_experiment}

% \begin{figure}[!t]
%     % \centering
%     \subfloat[]{\includegraphics[width=3.4in]{figures/UserCapacity_ber}}
    
%     \subfloat[]{\includegraphics[width=3.4in]{figures/UserCapacity}}
%     \caption{\red{exp} Evaluation of steganographic capacity. (a) BER. (b) PER. \red{maybe decrease the figure width? it's too huge now}}
%     \label{fig:steganographic_capacity}
% \end{figure}

\begin{figure}[!t]
    \centering
    \includegraphics[width=3.4in]{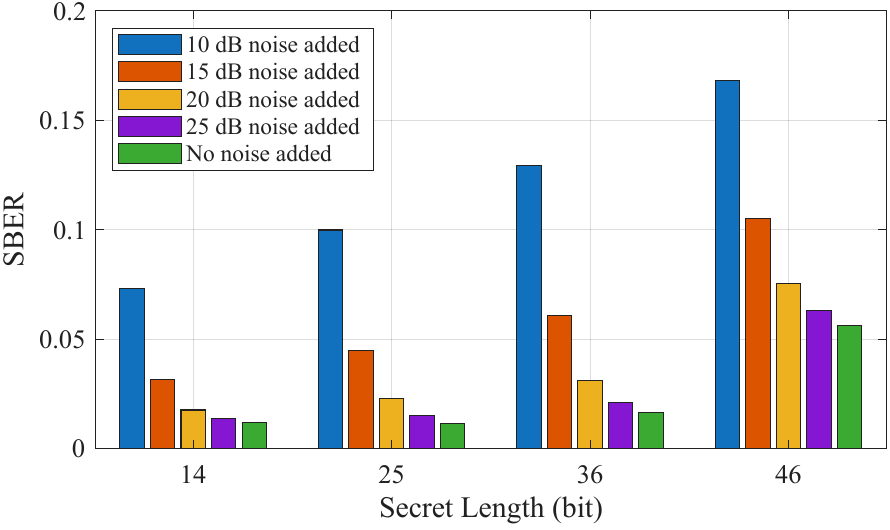}
    \caption{Evaluation of steganographic capacity.}
    \label{fig:steganographic_capacity}
\end{figure}

% \begin{figure}[!t]
%     \centering
%     \subfloat[]{\includegraphics[width=3.4in]{figures/experiment_ber.pdf}}
    
%     \subfloat[]{\includegraphics[width=3.4in]{figures/experiment_per.pdf}}
%     \caption{Evaluation of steganographic capacity. (a) BER. (b) PER.}
%     \label{fig:steganographic_capacity}
% \end{figure}

The simulation results presented in Section~\ref{sec:simulation_results} preliminarily evaluated the steganographic capacity. This is further evaluated through hardware prototypes.
We utilize two USRP B210 SDRs as the transmitter and receiver, respectively, placed approximately 1 meter apart in an indoor office environment. The length of the embedded secret is gradually increased from 14 bits to 46 bits, and the SBER is calculated. 
The received Wi-Fi signals consistently maintain an SNR above 30~dB. To emulate various SNR scenarios ranging from 10~dB to 25~dB, artificial noise is added to the received Wi-Fi baseband waveform represented by the raw I/Q samples.

% The experimental results are presented in Fig.~\ref{fig:steganographic_capacity}. It can be observed that both BER and PER increase with the length of the embedded secrets, exhibiting a trend consistent with the simulation results. Notably, when transmitting 46-bit secrets, the BER and PER are significantly higher than those observed in simulation, reaching 0.0647 and 0.7116, respectively. However, this performance degradation can be substantially improved through the implementation of more advanced FEC techniques rather than Hamming codes. In all other cases, both BER and PER remain below acceptable levels.
The experimental results are presented in Fig.~\ref{fig:steganographic_capacity}. It can be observed that SBER generally increases with the length of the embedded secrets, exhibiting a trend consistent with the simulation results. Nevertheless, the experimental SBER values are higher than their simulated counterparts, which can be attributed to real-world impairments, including hardware imperfections and environmental interference. Despite this performance degradation, the SBER remains within acceptable tolerance limits, i.e., less than 0.17 at 10~dB. The implementation of advanced forward error correction (FEC) techniques can effectively mitigate these transmission errors. For example, the Bose–Chaudhuri–Hocquenghem (BCH) codes can correct up to 25\% of errors under theoretical conditions~\cite{zhang2016efficient, al2017enhancing}.

In the proposed PHY information hiding scheme, each Wi-Fi packet carries 14 to 46 secret bits.
In the IEEE 802.11 standard, packet air times and inter-frame spaces are on the order of microseconds, so a Wi-Fi link can convey over a thousand packets per second.
The resulting secret throughput can therefore reach tens of kilobits per second when the link carries a dense packet stream; the per-packet embedding capacity itself remains limited and offers room for further improvement.

\subsection{Evaluation of Robustness to Channel Variations}

A robust PHY information hiding system must demonstrate resilience to channel variations, ensuring reliable secret recovery across diverse propagation environments.
To validate this requirement, we conducted extensive experiments in both indoor and outdoor environments. The secret length is fixed at 14 bits throughout this subsection. The experimental setup and results are detailed below.

\begin{figure}[!t]
    \centering
    \includegraphics[width=3.4in]{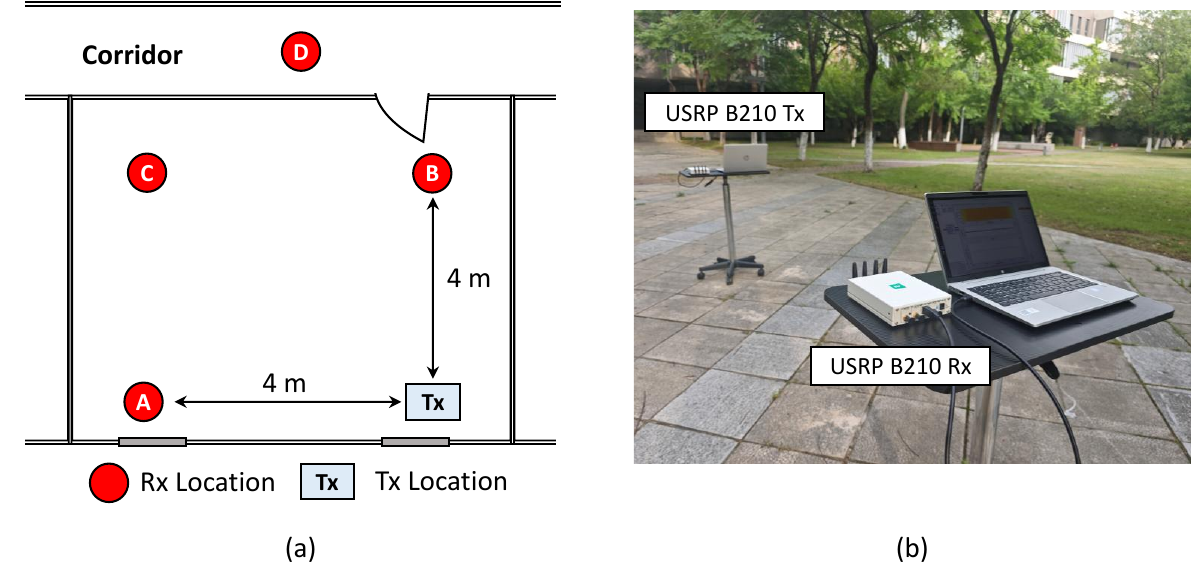}
    \caption{Experiment environments. (a) Indoor scenario in an office building. (b) Outdoor scenario in a park.}
    \label{fig:exp_environment}
\end{figure}

\begin{figure}[!t]
    \centering
    \includegraphics[width=3.4in]{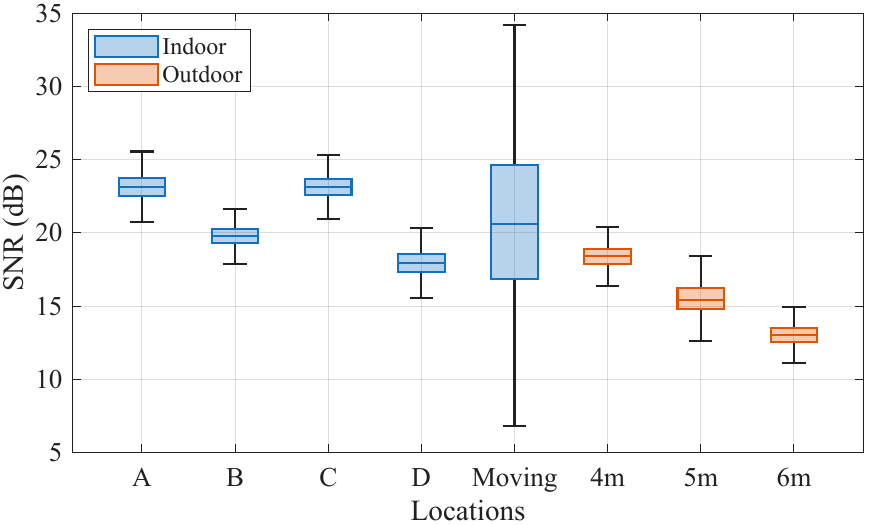}
    \caption{The estimated SNR values of indoor and outdoor scenarios.}
    \label{fig:snr_estimation}
\end{figure}

\begin{figure}[!t]
    \centering
    \includegraphics[width=3.4in]{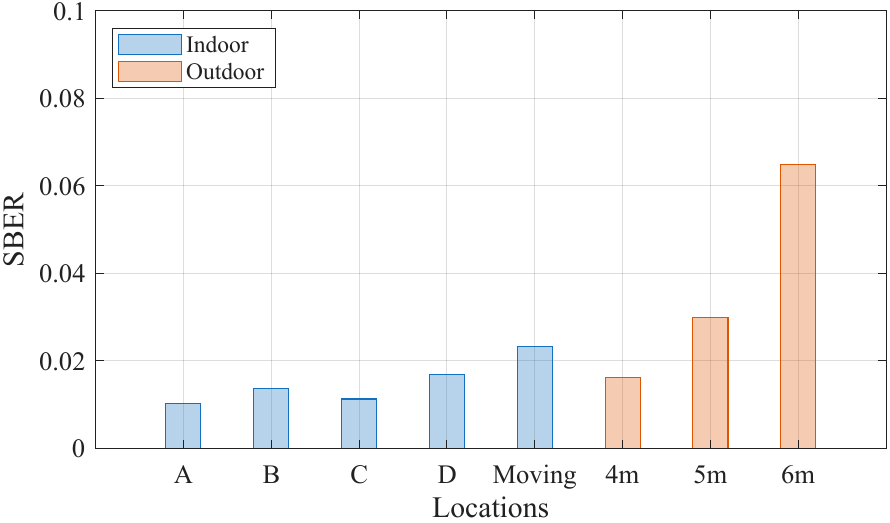}
    \caption{Experimental results of environmental robustness.}
    \label{fig:environmental_robustness}
\end{figure}

\textbf{Indoor Experiment:} Experiments were first conducted within an office building, with the floor plan shown in Fig.~\ref{fig:exp_environment}(a). The transmitter was positioned at the corner of the office, while the receiver was sequentially placed at four different locations, i.e., A, B, C, and D, to collect four distinct test datasets. Additionally, another dataset was recorded under a more challenging dynamic scenario, where the receiver moved randomly within the office to emulate realistic mobile usage conditions. The estimated SNR values corresponding to these five datasets are presented in Fig.~\ref{fig:snr_estimation}.

The evaluation results in the indoor experiments are presented in Fig.~\ref{fig:environmental_robustness}. The SBER remains below 0.02 across all datasets collected in static conditions, i.e., datasets A, B, C, D, indicating that the secrets can be accurately extracted at the receiver even if the receiver is relocated. Notably, the dataset collected under mobile conditions exhibited the highest SBER of 0.0233, which is expected given the increased channel variability and mobility-induced distortion in this setting.

\textbf{Outdoor Experiment:} We further conducted experiments in an outdoor park, with the surrounding environment depicted in Fig.~\ref{fig:exp_environment}(b). Pedestrians and vehicles moved randomly throughout the area. The USRP B210 transmitter was positioned at a fixed location, while the receiver was sequentially moved to distances of 4~m, 5~m, and 6~m to collect three distinct test datasets. The estimated SNR values are presented in Fig.~\ref{fig:snr_estimation}, which are generally lower than those collected in indoor scenarios.

The experimental results are presented in Fig.~\ref{fig:environmental_robustness}. The secret recovery performance remained satisfactory across all outdoor configurations. Even in the most challenging case with a transceiver separation of 6 m, the SBER was 0.065. The performance degradation is primarily attributed to the decrease in SNR, which dropped to approximately 13 dB at the 6 m distance. However, the performance can be further enhanced through the implementation of more advanced FEC schemes.

\begin{figure}[!t]
    \centering
    \includegraphics[width=3.4in]{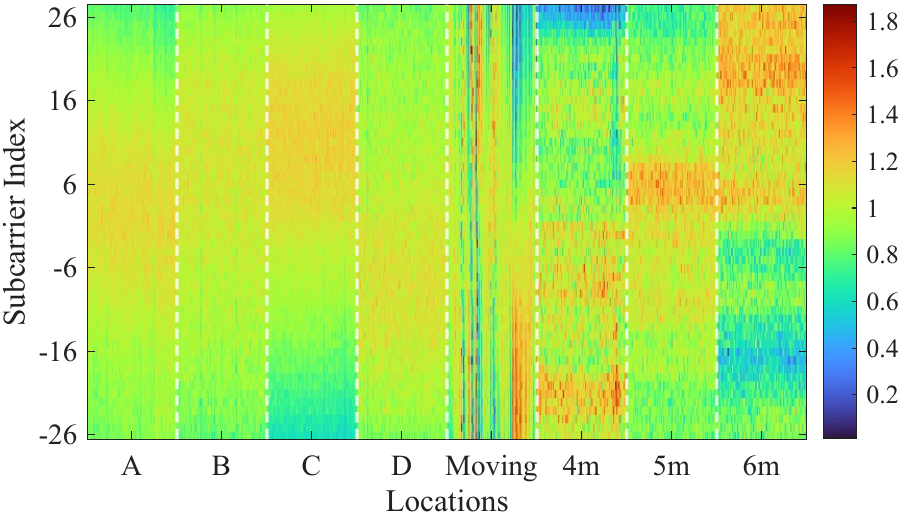}
    \caption{Magnitude of estimated CSI values in indoor and outdoor scenarios.}
    \label{fig:csi_estimation}
\end{figure}

To further illustrate the channel conditions encountered in our experiments, Fig.~\ref{fig:csi_estimation} visualizes the CSI magnitude across representative indoor and outdoor scenarios. To ensure clear visualization while maintaining statistical diversity, we display a subset of 100 packets per location. The results reveal the prevalence of frequency-selective fading induced by multipath propagation in both environments. Particularly noteworthy is the moving scenario, which exhibits rapid temporal fluctuations reflecting the dynamic channel evolution caused by receiver mobility. Despite these challenging propagation conditions, the consistently low SBER values demonstrated in Fig.~\ref{fig:environmental_robustness} confirm that our proposed method ensures robust secret recovery across diverse and time-varying wireless channels.

It should be noted that the relatively limited communication range was constrained by the low transmission power of the USRP B210 platform~\cite{usrpb210url}. Specifically, the SNR degraded to as low as 13~dB at the 6~m distance. The operational range could be significantly extended through the use of high-end hardware platforms with greater transmission power and enhanced receiver sensitivity.

\section{Discussion}\label{sec:discussion}

\subsection{Security Analysis}\label{sec:security_analysis}\label{sec:covertness}
\begin{figure}[!t]
    \centering
    \includegraphics[width=3.4in]{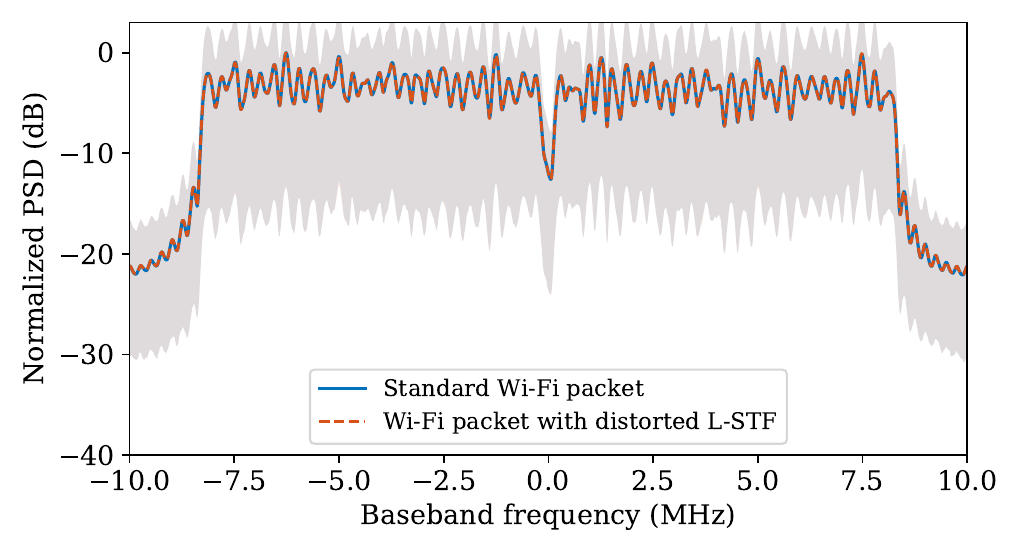}
    \caption{Complete-packet PSD of standard Wi-Fi packets and of packets carrying a secret in the L-STF, under TGn Model-B at 20~dB SNR. The shaded bands mark the per-frequency 5th--95th percentile range across packets. The grey region is where the two bands overlap.}
    \label{fig:psd_covertness}
\end{figure}

The security of the proposed PHY information-hiding scheme rests on two complementary properties. First, the secret message is embedded solely as a weak physical-layer perturbation of the L-STF waveform, rather than in the payload or any other decodable protocol field. Because this perturbation is deliberately kept imperceptible, the transmitted packet remains indistinguishable in appearance from a standard Wi-Fi waveform, rendering the embedded distortion difficult to detect. Fig.~\ref{fig:psd_covertness} quantifies this effect: the complete-packet PSD of packets carrying a secret in the L-STF coincides with that of standard Wi-Fi packets over both the occupied band and the out-of-band spectrum, so the embedding neither adds spectral support nor changes the spectral envelope. Moreover, commodity Wi-Fi chipsets expose no physical-layer interface: a conventional receiver built on such chipsets never obtains the raw I/Q samples of the preamble, and once synchronization completes, the preamble is discarded rather than forwarded to higher layers, which further enhances covertness. Second, even if an adversary becomes aware that a covert channel resides in the L-STF, merely capturing the over-the-air waveform does not suffice to recover the secret: successful extraction additionally requires the complete set of neural-network weights of the secret extractor shared privately between Alice and Bob.

\subsection{Complexity Analysis}\label{sec:complexity_analysis}
\begin{table}[!t]
    \centering
    \caption{Complexity of the designed neural networks.}
    \label{tab:complexity_analysis}
    \begin{tabular}{lrrr}
        \toprule
        Module & Parameters & Size (MB) & Latency (ms) \\
        \midrule
        Embedder  & 1,403,426 & 5.61 & 0.301 \\
        Extractor & 1,390,816 & 5.56 & 0.411 \\
        \bottomrule
    \end{tabular}
\end{table}
Table~\ref{tab:complexity_analysis} reports the parameter count, model size, and inference latency of the designed neural networks. The embedder contains 1,403,426 parameters, occupies 5.61~MB of FP32 storage, and runs in 0.301~ms per packet, while the extractor contains 1,390,816 parameters, occupies 5.56~MB, and runs in 0.411~ms per packet. These figures were obtained on the workstation described in Section~\ref{sec:nn_training_configuration}. Both networks are thus compact in terms of parameter count and model size.

\subsection{Compatibility Analysis}\label{sec:compatibility_analysis}

While our prototype is implemented over the 20~MHz SISO non-HT format, the proposed scheme is not confined to it.
The secret is embedded in the L-STF of the legacy preamble, a field that every subsequent IEEE 802.11 amendment retains at the start of the packet for backward compatibility. Consequently, IEEE 802.11n (HT), 802.11ac (VHT), 802.11ax (HE), and 802.11be (EHT) all begin with the same L-STF~\cite{IEEEstd}.
The proposed PHY information-hiding framework can therefore be directly applied to these formats or even future Wi-Fi standards without modification.
The MIMO case may require a different embedding strategy, with the secret hidden in the signals transmitted by the additional antennas.

\section{Conclusion}\label{sec:conclusion}
This paper presents a practical and robust PHY information hiding system for Wi-Fi, capable of embedding secrets within the outgoing PHY waveform without affecting the primary communication link. The secret embedding is achieved by intentionally introducing subtle and imperceptible distortion into the L-STF preamble using a pair of NNs. The original functionalities of L-STF remain unaffected since the periodic structure is preserved. This pair of NNs is trained in an end-to-end manner under emulated channel conditions, enabling reliable secret embedding and extraction. Simulation results confirm that the proposed PHY information hiding scheme does not compromise the performance of primary Wi-Fi links and demonstrate that the embedded secrets can be reliably extracted under various channel conditions. We further created a PHY information hiding prototype using SDR platforms and carried out experiments in both indoor and outdoor environments. Experimental results show that the system can embed 46-bit secrets with a worst-case SBER below 0.17 while preserving payload integrity across diverse channel conditions.

% \clearpage

% \section*{Acknowledgment}

% The work of Guanxiong Shen was supported in part by the National Natural Science Foundation of China under Grant 62401138. The work of Guanxiong Shen was supported in part by the National Natural Science Foundation of China under Grant 62401138.

% \clearpage
% \section{Notes}

% \red{Note 1: for upper-class lettered reference, use the command like this: journal={{IEEE} Std 802.11-2020 (Revision of {IEEE} Std 802.11-2016)}. include the IEEE with a pair of \{, \} }

% \red{Note 2: please specify the location to [!t] for all the figures and tables. }

% \red{Note 3: The second section, background and challenges, should include technical background information, and can incorporate formulas or charts if necessary. The current version looks more like part of the introduction.}

% \red{Note 4: `RF fingerprint' and `RF fingerprinting' are two distinct terminologies, I don't think `RFF' is a good abbrv., please use full name instead}

% \red{Note 5: typically we do not incorporate formulas in system overview, this (and the overview figure) is the first-read part for reviewers so please make it easy to follow.}

% \red{Note 6: why `packet detection' is a subsection of `Embedding of Digital RF fingerprint'? this is a receiver operation. I have moved it.}

% \bibliographystyle{IEEEtran}
% \bibliography{ref.bib}

% Bibliography inlined from compiled .bbl for arXiv submission.

\end{document}